\documentclass[a4paper,11pt]{article}
\pdfoutput=1 % if your are submitting a pdflatex (i.e. if you have
\usepackage{jcappub} % for details on the use of the package, please
\usepackage[utf8]{inputenc}
\usepackage[numbers,sort&compress]{natbib}

\newcommand{\be}{\begin{equation}}
\newcommand{\ee}{\end{equation}}
\newcommand{\bea}{\begin{eqnarray}}
\newcommand{\eea}{\end{eqnarray}}

\newcommand{\abs}[1]{\lvert#1\rvert}

\DeclareMathAlphabet{\mathcal}{OMS}{cmsy}{m}{n}

\title{Chaotic Inflation and Reheating in Generalized Scalar-Tensor Gravity}

\author[a]{Mat\'ias L\'opez,}
\author[a]{Giovanni Otalora}
\author[a]{and Nelson Videla }

\affiliation[a]{Instituto de F\'{\i}sica, Pontificia Universidad Cat\'olica de Valpara\'{\i}so,\\ Casilla 4950,  Valpara\'{\i}so, Chile \label{addr1}}

\emailAdd{matiasdamianlopezg@gmail.com}
\emailAdd{giovanni.otalora@pucv.cl}
\emailAdd{\\ nelson.videla@pucv.cl }

\abstract{In the present work, we study slow-roll inflation in scalar-tensor gravity theories in the presence of both the non-minimal coupling between the scalar field and curvature, and the Galileon self-interaction of the scalar field. Furthermore, we give predictions for the duration of reheating as well as for the reheating temperature after inflation. After working out the expressions for the power spectra of scalar and tensor perturbations in the case of a general non-minimal coupling function that depends solely on the scalar field and a general scalar potential, we focus on the special cases of the power-law coupling function and chaotic quadratic inflation. Thus, under the slow-roll approximation we confront the predictions of the model with the current PLANCK constraints on the spectral index $n_s$ and the tensor-to-scalar ratio $r$ using the $n_{s}-r$ plane. We found that the combination of the non-minimal coupling and Galileon self-interaction effects allows us to obtain better results for $r$ than in the case in which each effect is considered separately. Particularly, we obtained that the predictions of the model are in agreement with the current observational bounds on $n_{s}$ and $r$ within the $95 \%$ C.L region and also slightly inside the $68 \%$ C.L region. Also, we investigate the oscillatory regime after the end of inflation by solving the full background equations, and then we determine the upper bound for the Galileon and non-minimal coupling parameters under the condition that the scalar field oscillates coherently during reheating. Finally, after approximating reheating by a constant equation of state, we derive the relations between the reheating duration, the temperature at the end of reheating, its equation of state, and the number of $e$-folds of inflation and then we relate them all to the inflationary observables.}

\begin{document}
\maketitle
\flushbottom

%%%%%%%%%%%%%%%%%%%%%%%%%%%%%%%%%%%%%%%%%%%%%%%%%%%%%%%%%%%%%%%%%%%%
\section{Introduction}\label{Introduction}

Cosmic inflation \cite{starobinsky1,inflation1,inflation2,inflation3}-a stage of accelerated expansion of the very early universe-has been widely recognized as an elegant mechanism which solves the initial condition
problems of the standard hot Big Bang (HBB)
Model, for example, the flatness and horizon problems.  However, the essential feature of inflation
is that it generates primordial curvature perturbations, which seed cosmic microwave background (CMB)
temperature anisotropies and consequently the formation of Large Scale Structure (LSS) of the universe, and they are sourced by vacuum fluctuations of a scalar field, the so-called the inflaton \cite{Starobinsky:1979ty,R2,R202,R203,R204,R205,Abazajian:2013vfg}. Theoretically, in the simplest single-field slow-roll
inflation models, characterized by a minimal kinetic term and a potential $V(\phi)$, the spectrum of the primordial curvature perturbations is almost Gaussian
and almost scale-invariant according to the cosmological perturbation theory \cite{Malik:2008im,Chen:2006nt,Maldacena:2002vr,Riotto:2002yw,Wands:2000dp,Mukhanov:1990me,Bardeen:1980kt}. A further prediction of
inflationary models is the generation of tensor perturbations as a background of primordial gravitational waves (GWs), whose amplitude can be parametrized in terms of the the ratio
between the amplitude of the primordial tensor perturbation and that of the primordial
curvature perturbation called tensor-to-scalar ratio $r$ \cite{Lyth:1996im}.
As for the tensor-to-scalar ratio, we have not detected tensor perturbations until now.
The current observations, therefore, gives only upper bound on $r$. The PLANCK upper bound on the tensor-to-scalar-ratio \cite{planck2,planck4}, $r_{0.002}<0.10$ at 95$\%$ C.L., combined with the BICEP2/Keck Array (BK14) data \cite{bicep,Ade:2018gkx}, is further tightened, $r_{0.002}<0.064$.

The transition era after the end of inflation, during which the inflaton is converted into the particles that populate the Universe later on is called reheating \cite{reh1,reh2}, whose physics is complicated, highly uncertain, and in addition it cannot be directly probed by observations. For comprehensive reviews, see e.g. Refs. \cite{reh3,reh4,reh5}. One may obtain, however, indirect constraints on reheating according to the following strategy: First we parametrize our ignorance assuming for the fluid a constant equation of state (EoS) $w_{re}$ during reheating, and then we find certain relations between the reheating temperature $T_{re}$ and the duration of reheating $N_{re}$ with $w_{re}$ and inflationary observables \cite{Martin:2010kz,paper1,Martin:2014nya,paper2,Cai:2015soa,paper3,Rehagen:2015zma,Ueno:2016dim,Panotopoulos:2020qzi,Mishra:2021wkm,Osses:2021snt,Gong:2015qha}. On theoretical grounds, the temperature of reheating is assumed to be larger than the temperature of Electroweak (EW) transition \cite{reh4,reh5}. For some conservative issues, the temperature for the reheating era must be much larger by several orders of magnitude than the temperature reached in the big-bang nucleosynthesis (BBN), i.e., above a MeV \cite{bound1,bound2}.

One can test various models of
inflation, by comparing the theoretical predictions for the spectrum of the primordial 
perturbations with current observations, specially related with the CMB temperature anisotropies from the PLANCK collaboration \cite{planck2,planck4} as well as the BICEP2/Keck-Array data \cite{bicep,Ade:2018gkx}. Particularly, the constraints in the $n_s-r$ plane  give  us  the  predictions  of  a  number  of  representative inflationary potentials \cite{kolb}. In the most simplest models, so-called chaotic inflation \cite{inflation3}, 
the potential is chosen to be quadratic or quartic form, i.e. $\frac{m}{2}\phi^2$ or $\frac{\lambda}{4}\phi^4$, terms that are always present in the scalar potential of the Higgs sector in all renormalizable gauge field theories \cite{Pich:2007vu} in which the gauge symmetry is spontaneously broken via the Englert-Brout-Higgs mechanism \cite{Higgs:1964pj,Englert:1964et}. In despite their simplicity and  motivations from Particle Physics, the quadratic and quartic potentials, within the framework of General Relativity (GR), are in strong conflict with the recent combined analysis of PLANCK 2018 and BICEP2/Keck Array CMB, being ruling out since the predicted amount of tensor perturbations is too large when comparing with the current observed limit of $r_{0.002}<0.064$ at 95$\%$ C.L. due the trans-Planckian incursion of the inflaton field during inflation \cite{Lyth:1996im}. More generally, the monomial potential $V(\phi)=V_0 (\phi/M_{Pl})^n$ is ruled out by PLANCK 2018 data for $n\geq 2$ at 95$\%$ C.L. \cite{planck4}. 

From the theoretical and observational points of view, going beyond standard canonical inflation
within GR has become of a special interest. For instance, within the framework of quantum field theory in curved space-time, a non-minimal coupling (NMC) between the scalar field and curvature can naturally arise into the theory either by quantum corrections \cite{Linde:1982zj} or renormalizability requirements \cite{Freedman:1974gs,Freedman:1974ze,Birrell:1982ix}. What is more, in the cosmological context, a non-minimal coupling to gravity of the form $\xi \phi^2$ accounts in modifying the inflationary dynamics and bring chaotic inflation in better agreement with the current observational bounds \cite{Shokri:2019rfi,Tenkanen:2017jih,Boubekeur:2015xza,Linde:2011nh,Tsujikawa:2013ila,DeFelice:2011jm} when the coupling $\xi$ is large. Nevertheless, in the limit of large coupling, the very well known
problem of unitarity in the context of quantum field theory arises \cite{Burgess:2009ea,Barbon:2009ya}. On the other hand, non-canonical inflation yields a scalar propagation speed
$c_s$ and a large or small amount of non-Gaussianities \cite{DeFelice:2011uc,DeFelice:2013ar}. A special class of such a models,
dubbed k-inflation, non-linear function of the kinetic term or a coupling depending on
the scalar field $\phi$ with the kinetic term $X\equiv -\partial_{\mu}\phi\partial^{\mu}{\phi}/2$ are present in the Lagrangian \cite{Tsujikawa:2013ila,DeFelice:2011jm,ArmendarizPicon:1999rj,Garriga:1999vw}. Another scenario having a non-trivial structure may also be obtained by adding higher derivative quantum gravity corrections to the action such as for example a Galileon-type field self-coupling in the form $G(\phi,X) \Box \phi$, with $G$ being an arbitrary function of $\phi$ and $X$ \cite{Deffayet:2010qz}. The Dvali-Gabadadze-Porrati braneworld (DPG) model \cite{Dvali:2000hr} offers naturally a mechanism for a self-interaction of this kind, with $G\sim X$, through a non-linear interaction of the helicity-$0$ mode of the graviton \cite{Nicolis:2004qq}. More general functional forms of $G(\phi,X)$ have also been considered for instance in Refs. \cite{Deffayet:2010qz,Silva:2009km,Kobayashi:2009wr,Kobayashi:2010wa,DeFelice:2010jn,DeFelice:2010gb}. In principle, the Galileon self-coupling was introduced
as an infrared modification of GR in order to account the present acceleration
of the Universe without needing a cosmological constant. Regarding the inflationary universe, in Ref. \cite{DeFelice:2011jm} the authors have investigated the chaotic inflation in the context of general modified gravitational theories with non-minimal coupling term to curvature and Galileon self-interaction. Potential-driven Galileon inflation was also studied in \cite{Tsujikawa:2013ila,paperbase} for a Galileon-self coupling of the form $G(\phi,X)=-X/M^3$, putting chaotic inflation in agreement with current bounds on the tensor-to-scalar ratio available at that time \cite{wmap}. Currently, the latest data from the Keck Array/BICEP2 and PLANCK collaborations \cite{Ade:2018gkx} set stronger constraints
on the tensor the tensor-to-scalar ratio. Accordingly, a Galileon self-coupling of the form $G(\phi,X)\propto f(\phi)X$ was introduced in \cite{Ramirez:2018dxe}, reconciling the chaotic potential with current observations with a particular choice of $f(\phi)$. In \cite{Teimoori:2017jzo} it was studied G-inflation with a generalized expression for the Galileon self-interaction given by $G(\phi,X)\propto X^n$ (firstly proposed in \cite{DeFelice:2011jm}), while in \cite{Herrera:2018ker} the authors introduced a
further generalization, namely $G(\phi,X)\propto \phi^{\nu}\,X^n$. In both already mentioned works, it was found that the effect of the power $n$ is to lower the tensor-to-scalar ratio, while this effect
is enhanced when the Galileon self-coupling dominates over the standard kinetic term. The results obtained in Refs.\cite{Teimoori:2017jzo} and \cite{Herrera:2018ker} were used in \cite{Lopez:2019wmr} to explore the consistency of chaotic inflation with current bounds on $r$. However, it is very-well known the appearance  of instabilities if the Galileon term is still dominating over the standard kinetic term after the end of inflationary expansion, leading to a negative propagation
speed squared of scalar modes, yielding to the instability of small-scale
perturbations and spoiling the oscillations of the inflaton field during reheating phase \cite{paperbase}.

A novel and unifying framework is given by the so-called Horndeski theory \cite{horndeski}, which is the most general scalar-tensor theory having second order equations of motion. Interestingly, it includes both the canonical scalar field and k-essence \cite{mukhanov2}, while at the same time accommodates $f(R)$ theories, Brans-Dicke (BD) theory and Galileon gravity \cite{kobayashi,Kobayashi:2010cm,deffayet,charmousis}. The Horndeski theory has the attractiveness  of being applied for accounting the current accelerated expansion of the universe and the inflationary phase of the very early universe as well. Similarly, in the context of torsional modified gravity a Horndeski inspired construction has been proposed in Refs. \cite{Gonzalez-Espinoza:2019ajd, Bahamonde:2019shr} as well as a generalized scalar-torsion gravity theory proposed in Ref. \cite{Hohmann:2018rwf} and applied to inflation in Refs. \cite{Gonzalez-Espinoza:2019ajd,Gonzalez-Espinoza:2020azh,Gonzalez-Espinoza:2021qnv} and dark energy in Refs. \cite{ Gonzalez-Espinoza:2021mwr, Gonzalez-Espinoza:2020jss}. Regarding the observational constraints on the Horndeski theory, the nearly simultaneous detection of GWs GW170817 and the $\gamma$-ray burst GRB 170817A provides a tight constraint
on the propagation speed speed of GWs, $c_{\textup{GW}}$ \cite{TheLIGOScientific:2017qsa,Monitor:2017mdv}
\begin{equation}
-3\times 10^{-15}<c_{\textup{GW}}-1<7\times 10^{-16},
\end{equation}
which implies that GWs propagate at the speed of light. Consequently, in order to achieve $c_{\textup{GW}}=1$ irrespective of cosmological background within Horndeski's theory, its Lagrangian is restricted to be 
constructed only with non-minimal coupling, k-essence, and Galileon self-interaction terms \cite{Baker:2017hug,Sakstein:2017xjx,Langlois:2017dyl}.

In this way, the main goal of the present work is to investigate the dynamics of chaotic inflation within the framework of non-minimally coupled scalar-tensor gravity theory with Galileon self-interaction and 
its consistency with the stringent bounds
from CMB observations. Furthermore, we give predictions for the duration of reheating as well as for the reheating temperature after inflation.

We organize our work as follows. In Section \ref{GSTG} we briefly presents the dynamics of inflation within a generalized scalar-tensor gravity with non-minimal coupling to curvature and Galileon self-interaction of the scalar field. Furthermore in this section we obtain the expressions for the power spectra of scalar and tensor perturbations. In Section \ref{Chaotic_NMC_Galileon} we apply the results obtained in the previous section to the case of chaotic inflation (quadratic potential). In Section \ref{Osc_Reg} we study the oscillatory regime of the scalar field after the end of inflation in order to constrain the free parameters of the model. In section \ref{Reheating} we use the approximation of perfect fluid with a constant equation of state in order to study reheating after inflation. Finally, Section \ref{Concluding_Remarks} is devoted to the concluding remarks.

\section{Inflation in a generalized scalar-tensor gravity}\label{GSTG}

In  this  section  we  give  a  brief  review  on  the  background dynamics and the cosmological perturbations in the framework of inflation in a generalized scalar-tensor gravity.

\subsection{Cosmological background evolution}

Our starting point, is the action for the Galileon scenario with a non-minimal coupling to gravity $F(\phi)$, which becomes \cite{horndeski}
\be
S=\int{d^{4}x\sqrt{-g}\left[\frac{M_{pl}^2}{2}F(\phi) R+K(\phi,X)-G(\phi,X)\Box{\phi}\right]},
\label{Action1}
\ee where $X=-\partial_{\mu}{\phi}\partial^{\mu}{\phi}/2$, and $\Box{\phi}=\nabla_{\mu}\nabla^{\mu}{\phi}$ the Laplacian-Beltrami Operator.

For the function $K(\phi,X)$ and the Galileon self-interaction $G(\phi,X)$, we choose respectively \cite{paperbase}
\be
K(\phi,X)=X-V(\phi),
\ee and 
\be
G(\phi,X)=\frac{C}{M^3} X,
\label{Galileon}
\ee
where $C$ is a dimensionless constant to be fixed and $M$ is the Galileon mass scale.

By assuming a spatially flat Friedmann-Lema\^{i}tre-Robertson-Walker (FLRW) metric
 and a homogeneous scalar field $\phi=\phi(t)$,  then the modified Friedmann equations are found to be \cite{DeFelice:2011uc}
\bea
\label{H00}
&& 3 M_{Pl}^2 F H^2+3 M_{Pl}^2 H\dot{F}-X-V-\frac{6 C}{M^3} H \dot{\phi} X=0,\\
\label{Hii}
&& 3 M_{Pl}^2 F H^2+2 M_{Pl}^2 H \dot{F}+2 M_{Pl}^2 F\dot{H}+M_{Pl}^2 \ddot{F}+X-V-\frac{C}{M^3}\dot{\phi}^2\ddot{\phi}=0 ,\\
&& \left(1+\frac{6 C}{M^3} H \dot{\phi}\right) \ddot{\phi}+\left(3 H +\frac{9 C}{M^3} H^2 \dot{\phi}+\frac{3 C}{M^3} \dot{H} \dot{\phi}\right)\dot{\phi}+V_{,\phi}-6 M_{Pl}^2 H^2 F_{,\phi}-\nonumber\\
&& 3 M_{Pl}^2 \dot{H} F_{,\phi}=0.
\label{phi_Eq}
\eea 

We introduce the slow-roll parameters
\bea
&& \epsilon=-\frac{\dot{H}}{H^2},\:\:\: \delta_{\phi}=\frac{\ddot{\phi}}{H\dot{\phi}}\:\:\: \delta_{X}=\frac{X}{M_{Pl}^2 H^2 F},\nonumber\\
&& \delta_{F}=\frac{\dot{F}}{H F},\:\:\: \delta_{\dot{F}}=\frac{\ddot{F}}{H \dot{F}}, \:\:\: \delta_{G}=\frac{C\dot{\phi}X}{M_{Pl}^2 M^3 H F}.
\label{Slow_Para}
\eea Using these parameters and the background equations \eqref{H00} and \eqref{Hii}, one can write
\be
\epsilon=\delta_{X}-\frac{1}{2}\delta_{F}+3\delta_{G}+\frac{1}{2}\delta_{\dot{F}}\delta_{F}-\delta_{G}\delta_{\phi}.
\ee

Slow-roll inflation requires $\epsilon\ll 1$, and thus all the other parameters must also be small. Hence, under the slow-roll approximation we can write
\be
\epsilon=\delta_{X}-\frac{1}{2}\delta_{F}+3\delta_{G}+\mathcal{O}(\epsilon^2),
\label{epsilon}
\ee with $\delta_{X},\delta_{F},\delta_{G}\ll 1$.

Considering the slow-roll approximation, the Friedmann and Klein-Gordon (KG) equations reduce to
\bea
&& 3M_{Pl}^{2}H^{2}F\simeq V(\phi),\\
&& 3H\Dot{\phi}\left(1+\mathcal{A}\right) + V_{,\phi}-6 M_{Pl}^2 H^2 F_{,\phi} \simeq 0,
\label{phi_slow}
\eea
Respectively. We have defined $\mathcal{A}=\frac{3 C H\Dot{\phi}}{M^{3}}$ as the Galileon term for the field equation. It is pretty clear that if $M \rightarrow \infty$ and $\xi \rightarrow 0$ we stand over the standard case. 
These equations can also be written as
\bea
&& 3 H^2\simeq \frac{V}{M_{Pl}^2 F},\\
&& \frac{\dot{\phi}}{M_{Pl} H}\simeq \frac{F}{1+\mathcal{A}} \left[2 \frac{M_{Pl}F_{,\phi}}{F} -\frac{M_{Pl}V_{,\phi}}{V}\right].
\eea
From Eq. \eqref{phi_slow} we find
\be
\dot{\phi}(\phi)=\frac{M^3}{6 C H}\left[-1 +\sqrt{1+\frac{4 C V}{M^3}\left(2\frac{F_{,\phi}}{F}-\frac{V_{,\phi}}{V}\right)}\right],
\ee and therefore, 
\be
\mathcal{A}(\phi)=\frac{1}{2}\left[-1+\sqrt{1+\frac{4 C V}{M^3}\left(2\frac{F_{,\phi}}{F}-\frac{V_{,\phi}}{V}\right)}\right].
\ee
From Eq. \eqref{Q_s_1}, and in order to avoid that $Q_s$ becomes negative, we assume the condition $C\dot{\phi}>0$ \cite{paperbase}. For $C>0$ one has that $\dot{\phi}>0$ and then $2 F_{,\phi}/F>V_{,\phi}/V$. On the other hand, for $C<0$, it is satisfied $\dot{\phi}<0$ and therefore $2 F_{,\phi}/F<V_{,\phi}/V$. Without loss of generality we can take $C=\pm 1$. Also, as it is usually considered, the transition point from the Galileon-dominated regime to the standard inflation is calculated from the condition $ \mathcal{A}(\phi_{G})=1$ which leads us to 
\be
\left.\ \frac{V_{,\phi}}{V}\right|_{\phi=\phi_G}=-\left.\ \left(\frac{2 M^3}{CV}-\frac{2 F_{,\phi}}{F}\right) \right|_{\phi=\phi_G}.
\label{phi_G_1}
\ee

 Thus, using the definitions \eqref{Slow_Para}, we find
\be
\delta_{X}=\frac{X}{M_{Pl}^2 H^2 F}=\frac{1}{2 F}\left(\frac{\dot{\phi}}{M_{Pl} H}\right)^2, 
\ee
\be
\delta_{F}=\frac{\dot{F}}{H F}=\frac{M_{Pl} F_{,\phi}}{F} \left(\frac{\dot{\phi}}{M_{Pl} H}\right),
\ee

\be
\delta_{G}=\frac{C\dot{\phi}X}{M_{Pl}^2 M^3 H F}=\frac{\mathcal{A}}{6 F} \left(\frac{\dot{\phi}}{M_{Pl} H}\right)^2.
\ee

Hence, the equation \eqref{epsilon} yields
\begin{equation}
\epsilon=\left(\frac{F(\phi)}{1+\mathcal{A}(\phi)}\right) \left[2 \epsilon_{F}(\phi)-3 \sqrt{\epsilon_{F}(\phi) \epsilon_{V}(\phi)}+\epsilon_{V}(\phi)\right],
\label{epsilon_phi}
\end{equation}
where we have introduced the slow-roll parameters
\be
\epsilon_{V}(\phi)\equiv \frac{1}{2}\left(\frac{M_{Pl} V_{,\phi}}{V}\right)^2,\:\:\: \epsilon_{F}(\phi)\equiv \frac{1}{2}\left(\frac{M_{Pl} F_{,\phi}}{F}\right)^2.
\ee

Therefore, under the slow-roll approximation, if we divide the reduced Friedmann equation with the reduced Klein-Gordon equation, we can form the $Hdt$ term and then know the number of $e$-folds $N$ as
\bea
&& N\equiv \int_{t_{*}}^{t_{end}}{ H dt}=\int_{\phi_{*}}^{\phi_{end}}{\left(\frac{\dot{\phi}}{M_{Pl} H}\right)^{-1} \left(\frac{d\phi}{M_{Pl}}\right)},\nonumber \\
&& \simeq \int_{\phi_{end}}^{\phi_{*}}{\frac{1}{\sqrt{2}}\left(\frac{1+\mathcal{A}(\phi)}{F(\phi)}\right)\left[\frac{1}{\sqrt{\epsilon_{V}(\phi)}-2\sqrt{\epsilon_{F}(\phi)}}\right]\left(\frac{d\phi}{M_{Pl}}\right)}.
\label{phi_N}
\eea In this latter equation the field value at the end of inflation is calculated from the condition $\epsilon(\phi_{end})=1$ in \eqref{epsilon_phi}.

Below, we compute the second order action and then the inflationary observables. 

\subsection{Cosmological perturbations}

In order to study primordial fluctuations we start from perturbed metric
\be
ds^2=-\left(1+2\alpha\right)dt^2+2 \partial_{i}{\psi}dt dx^{i}+a^2\left[\left(1+2 \mathcal{R}\right)\delta_{i j}+h_{ij}\right]dx^{i} dx^{j},
\ee where $\alpha$, $\psi$, and $\mathcal{R}$ are scalar perturbations, while $h_{ij}$ are tensor perturbations which are transverse and traceless. Also, we use the uniform-field gauge, such that the perturbed scalar field, $\phi=\phi_{0}(t)+\delta{\phi}(t,x)$, satisfies $\delta{\phi}=0$ \cite{DeFelice:2011uc,Ohashi:2012wf}.

Thus, by expanding to second order the action \eqref{Action1}, and after using the Hamiltonian and momentum constraints, we obtain for the scalar perturbations the second order action \cite{DeFelice:2011uc}
\be
S=\int{dt d^{3}x a^3 Q_{s}\left[\dot{\mathcal{R}}^2-\frac{c_{s}^2}{a^2}\left(\partial \mathcal{R}\right)^2\right]},
\ee where 
\be
Q_{s}=\frac{w_{1}\left(4 w_{1} w_{3}+9 w_{2}^2\right)}{3 w_{2}^2},\:\:\: c_{s}^2=\frac{3 \left(2 w_{1}^2 w_{2} H-w_{2}^2 w_{1}+4 w_{1}\dot{w}_{1} w_{2}-2 w_{1}^2 \dot{w}_{2}\right)}{w_{1}\left(4 w_{1} w_{3}+9w_{2}^2\right)}, 
\ee with
\bea
&& w_{1}=M_{Pl}^2 F,\\
&& w_{2}= 2 M_{Pl}^2 H F-\frac{2 C}{M^3} X\dot{\phi}+M_{Pl}^2\dot{F},\\
&& w_{3}=-9 M_{Pl}^2 F H^2-9 M_{Pl}^2 H \dot{F}+3 X+\frac{36 C}{M^3}\dot{\phi} H X.
\eea As usual, the conditions $Q_{s}>0$ and $c_{s}^2>0$ are imposed in order to avoid ghosts and Laplacian instabilities, respectively. 

Now, we expand in terms of the slow-roll parameters the quantities $Q_{s}$ and $c_{s}^2$
\bea
\label{Q_s_1}
&& Q_{s}=\frac{M_{Pl}^2 F \left[3 (\frac{1}{2}\delta_{F}-\delta_{G})^2+6\delta_{G}+\delta_{X}\right]}{(\frac{1}{2}\delta_{F}-\delta_{G}+1)^2}\simeq M_{Pl}^2 F \left(\delta_{X}+6 \delta_{G}\right),\\
&& c_{s}^2=\frac{\frac{1}{2}\delta_{F} (1-\delta_{\dot{F}}-2 \delta_{G})+\frac{3}{4} \delta_{F}^2+\delta_{G} (1-\delta_{G}+3 \delta_{\phi})+\epsilon}{3  (\frac{1}{2}\delta_{F}-\delta_{G})^2+6 \delta_{G}+\delta_{X}}\simeq \frac{\delta_{X}+4 \delta_{G}}{\delta_{X}+6 \delta_{G}}\label{cs2g}.
\eea

Also, we define the parameter
\be
\epsilon_{s}\equiv \frac{Q_{s}c_{s}^2}{M_{Pl}^2 F}\simeq \delta_{X}+4\delta_{G}.
\ee

After calculating the two-point correlation function of the curvature perturbation through the standard method
of quantizing the fields on a quasi de Sitter background and by using the solution for $\mathcal{R}$ obtained from the Mukhanov-Sasaki equation, we obtain the scalar power spectrum
\bea
&& \mathcal{P}_{s}=\left.{\frac{H^2}{8 \pi^2 Q_{s} c_{s}^3}}\right|_{c_{s}k=aH}=\left.{\frac{H^2}{8 \pi^2 M_{Pl}^2 F c_{s} \epsilon_{s}}}\right|_{c_{s}k=aH},\\
&& \simeq \frac{H^2 \left(\delta_{X}+6 \delta_{G}\right)^{\frac{1}{2}}}{8 \pi^2 M_{Pl}^2 F\left(\delta_{X}+4 \delta_{G}\right)^{3/2}}.
\eea

Thus, the spectral index is 
\be
n_{s}-1\equiv \left.{\frac{d\ln \mathcal{P}_{s}}{d \ln k}}\right|_{c_{s}k=aH}\simeq -2\epsilon-\delta_{F}-\eta_{s}-s.
\ee where we defined
\be
\eta_{s}\equiv \frac{\dot{\epsilon}_{s}}{H \epsilon_{s}},\:\:\:\ s=\frac{\dot{c}_{s}}{H c_{s}},
\ee and we assumed that $c_{s}^2$ is a slowly varying function with $s\ll 1$. 

For tensor perturbations we use the decomposition in the polarization modes in the form $h_{ij}=h_{+} e_{i j}^{+}+h_{\times} e_{i j}^{\times}$, which gives the second order action
\be
S_{T}^{(2)}=\sum_{p}{\int{dt dx^3 a^3 Q_{T}\left[\dot{h}_{p}^2-\frac{c_{T}^2}{a^2}\left(\partial h\right)^2\right]}}, 
\ee being $p=+,\times$, and 
\be
Q_{T}=\frac{w_{1}}{4},\:\:\:\:\: c_{T}^2=1,
\ee Thus, the above action describes the usual transverse massless graviton modes, propagating at speed of light, and then there is not a presence of any Laplacian instability. The condition to avoid ghost instabilities is $Q_{T}>0$. Following the same procedure as for the scalar sector, the tensor power spectrum is given by 
\be
\mathcal{P}_{T}=\left.\frac{H^2}{2 \pi^2 Q_{T} c_{T}^3}\right|_{c_{s} k=a H},
\label{P_t}
\ee which is also evaluated at the moment of the Hubble horizon crossing.

The tensor spectral index is 
\be
n_{T}\equiv \left.{\frac{d \ln \mathcal{P}_{T}}{d \ln k}}\right|_{c_{s}k=aH}=-2 \epsilon-\delta_{F}.
\ee

Then the tensor-to-scalar ratio is

\be
r=\frac{\mathcal{P}_{T}}{\mathcal{P}_{s}}=16 c_{s} \epsilon_{s},
\label{rr}
\ee which is written as 
\be
r\simeq 8 c_{s}\left(-n_{T}+2 \delta_{G}\right),
\ee
so it shows that the tensor spectral index and the tensor-to-scalar ratio are not independent parameters. 

From Eq. (\ref{rr}), it can be seen that inflation within the framework
of generalized scalar-tensor gravity becomes distinguishable, at phenomenological level, from standard inflation, for which $c^2_s=1$. Regarding higher order statistics of perturbations, if primordial
non-Gaussianities (NG) are eventually detected in future observations, by means measuring the non-linear parameter $f_{NL}$, it could break the degeneracy among the several inflation models
and also accounts in discriminating
between single-field inflation and other alternative scenarios (for a comprehensive review see, Refs.~\cite{Bartolo:2004if,Renaux-Petel:2015bja}). Particularly, for the simplest models of inflation, consisting in a single-field with a canonical kinetic term and
a smooth potential, the predicted amount of NG is such that $f_{NL}\ll 1$ \cite{Maldacena:2002vr}. 
Interestingly, a large amount of NG, $|f_{NL}|\gg 1$, arises by going further the previous properties. 
For instance, for models with $c_s^2\ll 1$, the non-linear parameter can be as large as
$|f_{NL}|> 1$ \cite{DeFelice:2011uc,DeFelice:2013ar}. If inflation is driven by both a non-minimal coupling $F(\phi)$ and a Galileon
self-coupling $G=-\frac{X}{M^3}$, from Eq.~(\ref{cs2g}), the scalar propagation speed squared
can be written as
\begin{equation}
c_{s}^2\simeq\frac{1+4\,\delta_{G}/\delta_{X}}{1+6\,\delta_{G}/\delta_{X}}\label{cs2r}.
\end{equation}
We note that in the regime $\delta_G\gg \delta_X$, the scalar propagation speed squared
reduces to $c^2_s\simeq 2/3$ which is not much smaller than 1 and then $\abs{f_{NL}}< 1$ \cite{DeFelice:2011jm,DeFelice:2011uc}. Furthermore, in the regime where $\delta_G$ and $\delta_X$ are of 
the same order, the scalar propagation speed squared becomes $c_s^2\simeq 5/7$, implying
a smaller value for $f_{NL}$ according to \cite{DeFelice:2011uc,DeFelice:2013ar}.

Current observational constraints on $f_{NL}$ are found to be $\abs{f_{NL}}\lesssim {\mathcal{O}}(10)$ \cite{Planck:2015zfm}. In particular, by combining temperature and polarization data, Planck collaboration has found the following constraints on $f_{NL}$ for the local, equilateral, orthogonal, and enfolded configurations  \cite{Planck:2015zfm}
\begin{eqnarray}
f_{NL}^{\textup{local}}&=&0.8\pm 5.0\,\,\,\,\,\,(68\%\,\textup{CL}),\\
f_{NL}^{\textup{equil}}&=&-4\pm 43\,\,\,\,\,\,\,\,(68\%\,\textup{CL}),\\
f_{NL}^{\textup{ortho}}&=&-26\pm 21\,\,\,\,\,(68\%\,\textup{CL}),\\
f_{NL}^{\textup{enfold}}&=&11\pm 32\,\,\,\,\,\,\,\,\,(68\%\,\textup{CL}).
\end{eqnarray}

In this way, if future observations
confirm the value $|f_{NL}|> 1$ at more than 95$\%$ CL, our inflationary model would be ruled out, and then we will have to go beyond the slow-variation single-field inflationary scenario.

\section{Chaotic inflation with non-minimal coupling and Galileon self-interaction}\label{Chaotic_NMC_Galileon}

We introduce the quadratic potential
\be
V(x)=V_{0} x^2,
\label{V_x}
\ee and the non-minimal coupling function
\be
F(x)=1+\frac{\xi}{2} x^2,
\label{F_x}
\ee where $V_{0}$, and $\xi$ are constants and we defined $x\equiv \phi/M_{Pl}$.
In this case one obtains
\be
\mathcal{A}(x)= \frac{1}{2} \left[-1+\sqrt{1+8 \delta  C x \left(1-\frac{4}{\xi  x^2+2}\right)}\right],
\ee where we defined $\delta=V_{0}/(M_{Pl} M^3)$.

Also, for these functions we get
\be
\epsilon_{V}(x)=\frac{2}{x^2},
\ee and
\be
\epsilon_{F}(x)=\frac{2 \xi ^2 x^2}{\left(\xi x^2+2\right)^2}.
\ee
Thus, from Eq. \eqref{epsilon_phi}, we find
\be
\epsilon(x)=\frac{2\left(2- \xi  x^2\right)}{x^2 (\mathcal{A}(x)+1) \left(\xi  x^2+2\right)}.
\label{epsilon_x}
\ee The value of the field at the end of inflation is calculated from the condition $\epsilon(x_{end})=1$. However, this equation cannot be solved analytically, and therefore we do it numerically.

 The transition point $\phi=\phi_{G}$ can be obtained using Eq. \eqref{phi_G_1}, which now takes the form
\be
\frac{1}{x_{G}}=\frac{\xi  x_{G}}{\left(\frac{1}{2} \xi  x_{G}^2+1\right)}-\frac{1}{\delta  C x_{G}^2},
\label{Trans_xG}
\ee where $x_{G}=\phi_{G}/M_{Pl}$. Thus, after solving the latter equation for $x_{G}=x_{G}(\delta,\xi)$, a function of $\delta$ and $\xi$, the condition $\epsilon(x_{G})\leq 1$ yields
\be
\frac{2- \xi  x_{G}(\delta,\xi)^2}{x_{G}(\delta,\xi)^2\left(\xi  x_{G}(\delta,\xi)^2+2\right)}\leq 1,
\label{delta_c}
\ee being that the equality is satisfied for the critical value of the Galileon parameter $\delta=\delta_{c}(\xi)$ which is a function of the non-minimal coupling parameter $\xi$. For $\delta<\delta_{c}(\xi)$ the Galileon-dominated regime ends before the end of slow-roll inflation, while for $\delta>\delta_{c}(\xi)$ the dominance of the Galileon term extends until after the end of inflation. Let us notice that once we constrain the parameter $\xi$ from the $n_{s}-r$ plane it will be possible to obtain the values of function $\delta_{c}(\xi)$. This is different to what happens in the case of a minimal coupling to gravity where $\delta_{c}$ is a fixed number depending only on the properties of the scalar potential \cite{Ohashi:2012wf}.  

The number of $e$-folds $N$ is obtained from \eqref{phi_N} which gives
\be
N\simeq \int_{x_{end}}^{x_{*}}{\frac{x (\mathcal{A}(x)+1)}{2-\xi  x^2} dx}.
\ee From this latter equation and after performing the numerical integration we can solve $x^{*}\equiv x(N)$ in terms of the $e$-folds number $N$. So, for our integration we use the initial conditions at the moment of the horizon crossing and the numerical solution for $x_{end}$ from Eq. \eqref{epsilon_x}.

The scalar power spectrum becomes 
\be
\mathcal{P}_{s}\simeq \frac{\sqrt{3} V_{0} x^4 (\mathcal{A}(x)+1)^2 \sqrt{2 \mathcal{A}(x)+1} \left(\xi  x^2+2\right)}{2 \pi ^2 M_{Pl}^4 (4 \mathcal{A}(x)+3)^{3/2} \left(\xi ^2 x^4-4\right)^2}.
\label{Ps_x}
\ee 

Also, the spectral index is written as
\bea
&& n_{s}-1\simeq \frac{3 \left(\xi ^2 x^4+8 \xi  x^2-4\right)}{2 x^2 (2 \mathcal{A}(x)+1) \left(\xi  x^2+2\right)}-\frac{\xi ^2 x^4+8 \xi  x^2-4}{2 x^2 (2 \mathcal{A}(x)+1)^2 \left(\xi  x^2+2\right)}-\nonumber\\
&& \frac{9 \left(\xi ^2 x^4+8 \xi  x^2-4\right)}{x^2 (4 \mathcal{A}(x)+3) \left(\xi  x^2+2\right)}-\frac{12 \left(2-\xi  x^2\right)}{x^2 (\mathcal{A}(x)+1) \left(\xi  x^2+2\right)},
\eea and the tensor-to-scalar ratio is given by
\bea
&& r \simeq \frac{16 (4 \mathcal{A}(x)+3)^{3/2} \left(2-\xi  x^2\right)^2}{3 x^2 (\mathcal{A}(x)+1)^2 \sqrt{6 \mathcal{A}(x)+3} \left(\xi  x^2+2\right)}.
\eea
From the above expressions for $n_{s}$ and $r$ evaluated at value of the field at the horizon crossing we can compare the theoretical predictions for the model with the current observational constraints. Therefore, after constraining the parameters $\xi$ and $\delta$ from the $n_{s}$-$r$ plane, and by using Eq. \eqref{Ps_x} with the current observational value for the amplitude of primordial scalar perturbations $\mathcal{P}_{s}=2.141\times 10^{-9} $ \cite{planck4}, we can estimate the ranges for $V_{0}$ and then for the Galileon mass term $M$. 

\begin{figure}[htbp]
	\centering
		\includegraphics[width=0.6\textwidth]{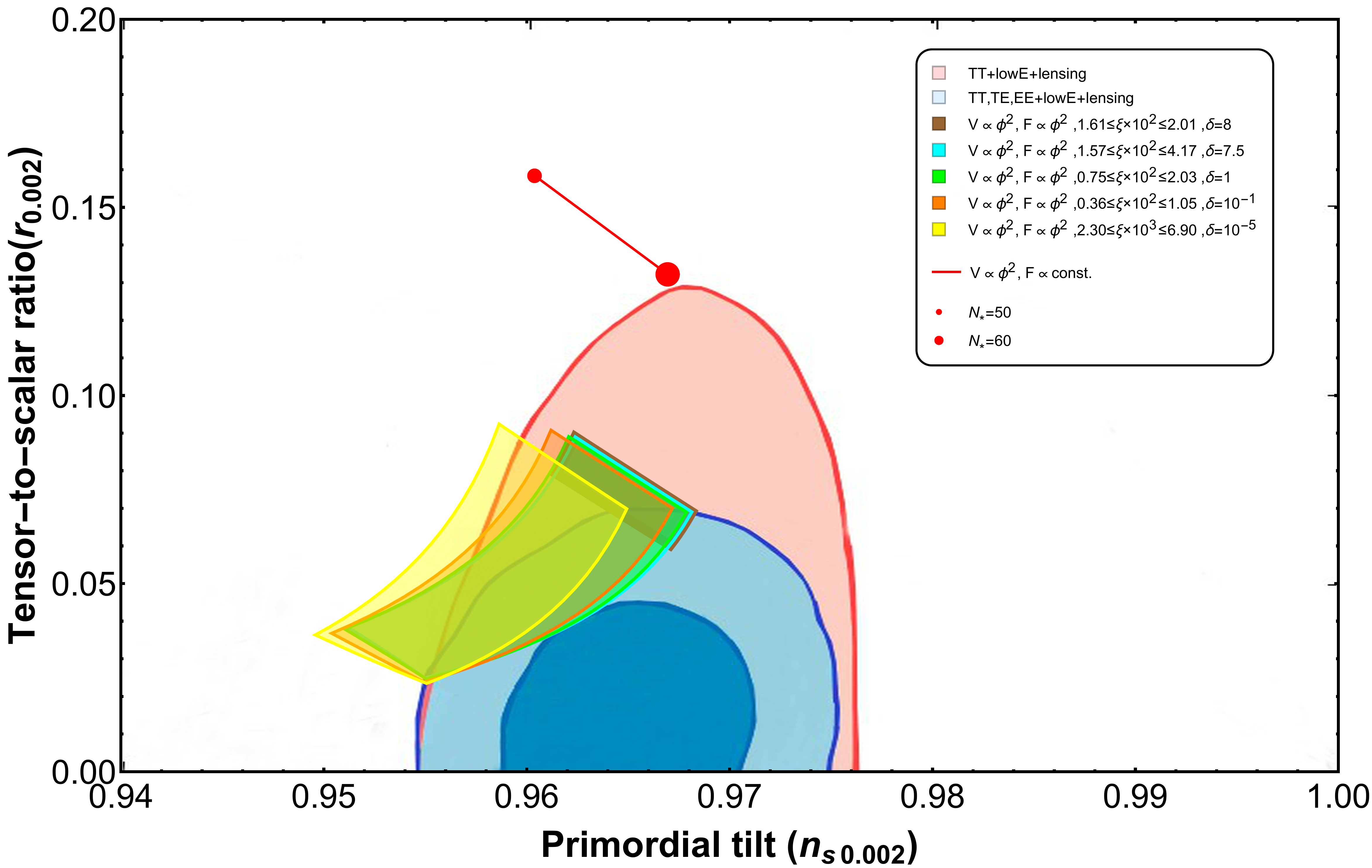}
	\caption{We depict the $(n_{s},r)$ plane for chaotic inflation in scalar-tensor gravity in the presence of both a non-minimal coupling to curvature and Galileon self-interaction term. For $N=60$, we find the corresponding ranges of the non-minimal coupling parameter, $\xi$, for each different value of the parameter related to the Galileon self-interaction, $\delta$, in which the predictions of the model are in agreement with the $95\%$ C.L region of PLANCK $2018$ data \cite{planck4}. For the highest value of $\delta$ and $\xi$ we have used the constraints found from the study of the oscillatory regime of the inflaton after the end of inflation.}
	\label{r_ns_compressed}
\end{figure}

In FIG. \ref{r_ns_compressed} we depict the $n_{s}-r$ plane with the the marginalized joint $68 \%$ and $95 \%$ CL regions at $k=0.002$ $\text{Mpc}^{-1}$ from PLANCK data \cite{planck4}, along with the theoretical predictions for chaotic inflation (quadratic potential) in generalized scalar-tensor gravity with Galileon self-interaction term. It is found that the predictions of the model are within the $95 \%$ CL region of PLANCK. Particularly, for $N=60$, and several different values of the $\delta$ parameter we have found the corresponding ranges for the non-minimal coupling parameter $\xi$, and then the ranges for $V_{0}$ and $M$. Thus, we put the all the physical results for the model in Table \ref{Table1} including the constraints obtained on the mass scale $m_{\phi}$ of the inflaton field and the Galileon mass $M$. Furthermore, we have numerically calculated the function $\delta_{c}(\xi)$ for the critical value of the Galileon parameter, such that we have found
\be
1.011 \lesssim \delta_{c}(\xi)\lesssim 1.031 \:\:\: \text{for}\:\:\: 7.5\times 10^{-3}\lesssim \xi\lesssim 2.05\times 10^{-2},
\ee  and the values for $n_{s}$ and $r$ within the $95 \%$ CL region of PLANCK.

One can observe from the results obtained in FIG. \ref{r_ns_compressed} that the parameter of the Galileon self-interaction $\delta$ plays an important role in lowering the predictions for tensor-to-scalar ratio parameter $r$ in non-minimally coupled scalar-tensor theories. However, using only the results from $n_{s}-r $ plane is not enough to obtain an upper bound for $\delta$, or equivalently, the lower bound for the Galileon mass $M$. Furthermore, it is well known that very large values of the Galileon self-interaction during slow-roll inflation may to disable the oscillatory regime of the inflaton after the end of inflation and then spoiling reheating \cite{Ohashi:2012wf}. Thus, taking into account that this problem is accompanied by the appearance of a negative propagation speed squared $c_{s}^2$ of the scalar mode, below we study the oscillations of the field after the end of inflation in order to find the upper bound for the parameter $\delta$.

\def\tablename{Table}%
\begin{table*}[!ht]
\centering
\begin{center}
\renewcommand{\tabcolsep}{1.0pc} % enlarge column spacing
\renewcommand{\arraystretch}{1} % enlarge line spacing
\begin{tabular}{| c | c | c | c | c |}\hline\cline{1-5}
$\delta$ &   $\xi \times 10^{2}$  & $(V_{0}/M_{Pl}^4)\times 10^{10}$  & $ (m_{\phi}/M_{Pl}) \times 10^6$   & $(M/M_{Pl})\times 10^{4}$ 
\\\cline{1-5} 
 $10^{-5}$ & $(0.230, 0.690)$ & $(0.162, 0.111)$ & $(5.69, 4.71)$ &  $(117, 104)$ \\\cline{1-5}
 $10^{-1}$ & $(0.360, 1.05)$ & $(0.264, 0.171)$ & $(7.27, 5.84)$ &  $(6.42, 5.55)$ \\\cline{1-5} 
 $1$ & $(0.750, 2.03)$ & $(0.531, 0.340)$ & $(10.3, 8.25)$ &  $(3.76, 3.24)$ \\\cline{1-5} 
 $7.5$ & $(1.57, 4.17)$  & $(1.10, 0.705)$ &  $(14.9, 11.9)$ &   $(2.45, 2.11)$\\\cline{1-5}
 $8$ & $(1.61, 2.01)$ & $(1.13, 1.09)$ & $(15.0, 14.8)$ & $(2.42, 2.39)$\\\cline{1-5}
 \hline
\end{tabular}
\end{center}
\caption{Summary on the parameters $\xi$, $V_{0}$, and $M$, for some values of the parameter $\delta$ for chaotic inflation (quadratic potential) and the number of $e$-folds of inflation $N=60$ in generalized scalar-tensor gravity with Galileon self-interaction term. Also, for the highest value of $\delta$ and $\xi$ we have used the constraints found from the study of the oscillatory regime after inflation. } 
\label{Table1}
\end{table*}

\section{Oscillatory regime of the inflaton}\label{Osc_Reg}

Let us introduce the following dimensionless variables in order to write in a convenient way the complete set 
of background equations
\bea
&& x=\frac{\phi}{M_{Pl}},\:\:\:\: y=\frac{\dot{\phi}}{M M_{Pl}},\:\:\:\: z=\frac{H}{M},\nonumber\\
&& U(x)=\frac{V(x)}{M^2 M_{Pl}^2},\:\:\:\: \beta =\frac{M_{Pl}}{M},
\eea  for the which the constraint equation \eqref{H00} becomes
\be
\frac{3 z^2 F}{\beta^2}=-\frac{3 y z F_{,x}}{\beta^2}+\frac{U}{\beta^2}+\frac{y^2}{2 \beta^2}-\frac{3 y^3 z}{\beta }.
\label{Constraint_eq}
\ee

Furthermore, after combining Eqs. \eqref{H00} and \eqref{Hii} to eliminate $V$, and using Eq. \eqref{Hii} and  \eqref{phi_Eq}
to solve $\ddot{\phi}$ and $\dot{H}$, we obtain
\bea
\label{x_N}
&& \frac{dx}{dN}=\frac{y}{z},\\
\label{y_N}
&& \frac{dy}{dN}=\Bigg[3 \beta  y^4 \left(F_{,xx}+1\right)+2 F \left(U_{,x}-6 z^2 F_{,x}\right)-12 \beta  y^3 z F_{,x}+\nonumber\\
&& 3 y z \left(2 F-F_{,x}^2\right)+3 y^2 \left(F_{,x} \left(F_{,xx}+1\right)-6 \beta  z^2 F\right)-9 \beta ^2 y^5 z\Bigg]/\nonumber\\
&& \Bigg[z \left(2 F (6 \beta  y z-1)-3 \left(F_{,x}+\beta  y^2\right)^2\right)\Bigg],\\
&& \frac{dz}{dN}=\Bigg[-6 \beta  y^3 z \left(F_{,xx}+2\right)+F_{,x} \left(6 z^2 F_{,x}-U_{,x}\right)-4 y z F_{,x}+27 \beta^2 y^4 z^2+\nonumber\\
&& y^2 \left(F_{,xx}+21 \beta  z^2 F_{,x}-\beta  U_{,x}+1\right)\Bigg]/\Bigg[z \left(2 F (6 \beta  y z-1)
-3 \left(F_{,x}+\beta  y^2\right)^2\right)\Bigg].
\label{z_N}
\eea In the above equations we introduced the $e$-folds number $N\equiv \log{a}$. Also, by using the constraint equation \eqref{Constraint_eq} we can eliminate Eq. \eqref{z_N}, and then we are left with only Eqs. \eqref{x_N} and \eqref{y_N}. 

We are interested in numerically solving the above system of equations for the chaotic potential (quadratic potential) in Eq. \eqref{V_x} and the non-minimal coupling function in Eq. \eqref{F_x}, in order to study the oscillatory regime of the scalar field after the end inflation. We choose the initial conditions at the horizon crossing and evolve the functions $x(N)$ and $y(N)$ until some $e$-folds after inflation. Thus, when fixing the initial conditions we can use the slow-roll approximation to find them, and also one can see that $\beta=\left(M_{Pl}^4 \delta/V_{0}\right)^{1/3}$ with $V_{0}$ calculated from Eq. \eqref{Ps_x}. Let us note that since the slow-approximation is less accurate at the end of inflation, after doing the estimation of the initial conditions, we also need to adjust better them in order to have the exact value of $60$ $e$-folds at the end of inflation.

\begin{figure}[htbp]
	\centering
		\includegraphics[width=0.60\textwidth]{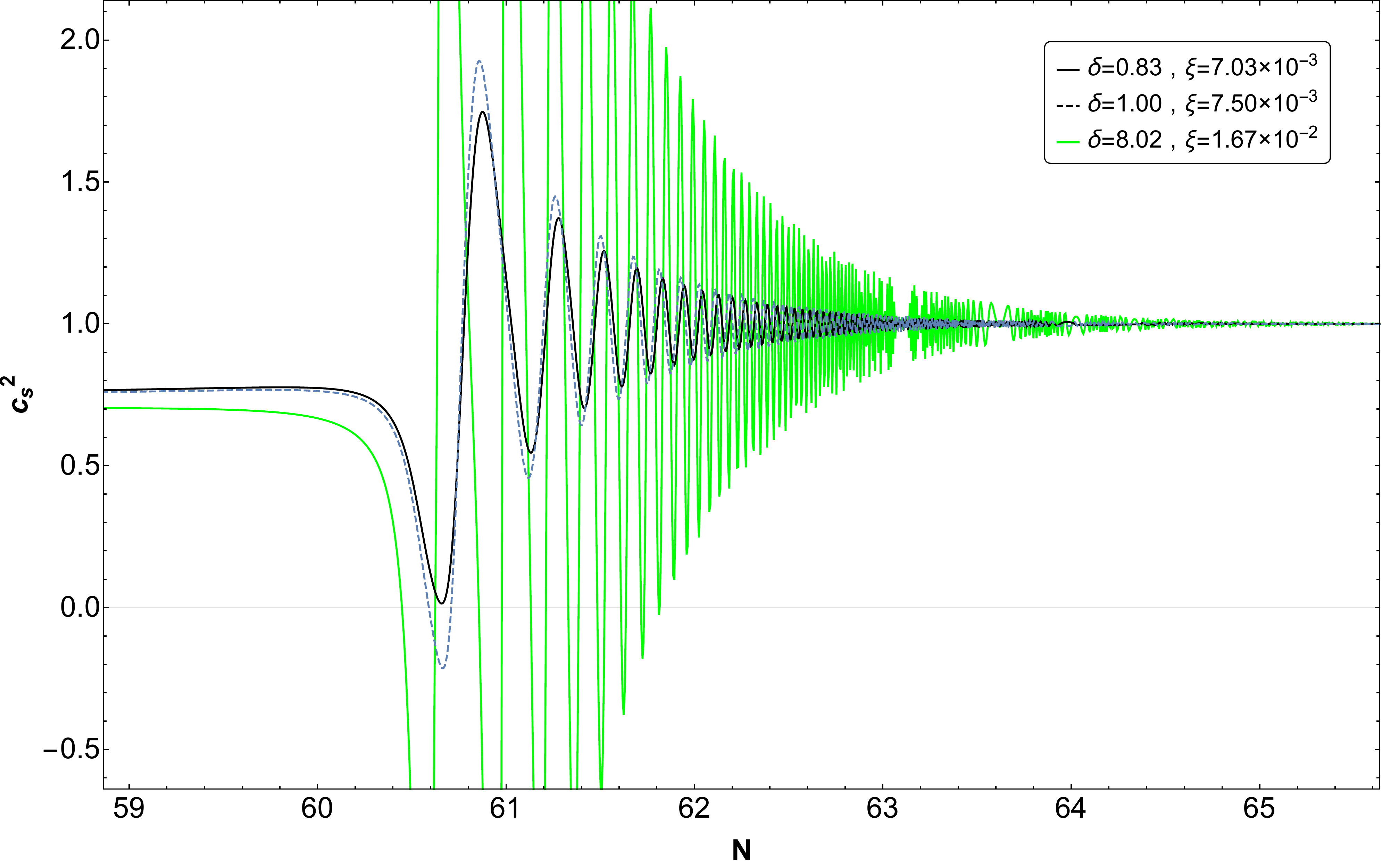}
	\caption{\small{Plot of the squared speed of scalar perturbations, for the quadratic chaotic potential and Galileon self-interactive term $G=-X/M^{3}$. Three different cases are represented. The values of our free parameters were restricted in the $n_{s}$-$r$ plane, and the initial condition for the inflaton field arise from solving the field equations in the slow-roll limit at $N=60$. These initial conditions were chosen in order to reach $\epsilon=1$ at the end of inflation. They are: $\phi^{*}=8.0968 \, M_{Pl}$ for the black line, $\phi^{*}=7.8433 \, M_{Pl}$ for the blue dashed line and $\phi^{*}=5.3757 \, M_{Pl}$ for the green one.  }}
	\label{cs2}
\end{figure}

\begin{figure}[htbp]
	\centering
		\includegraphics[width=0.50\textwidth]{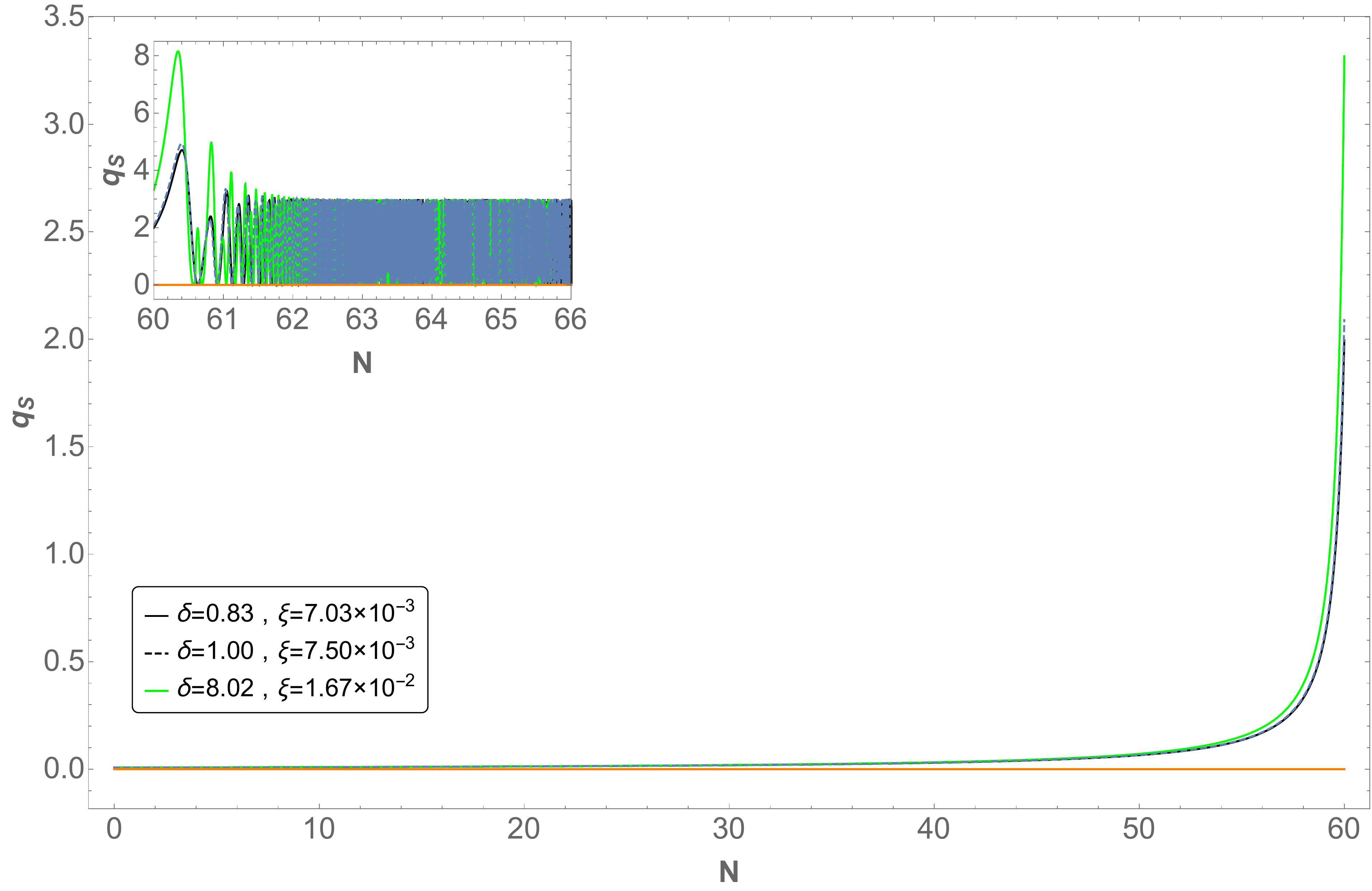}
	\caption{We show that the parameter $q_{s}\equiv Q_{s}/M_{Pl}^2 F $ is always positive. In particular, if $\delta=0.83$ and $\xi\simeq 7.035 \times 10^{-3}$, the $q_{s}(N)$ function will reach a maximum peak with $q_{s}\simeq 5$ just after the end of inflation and then it will oscillate
	always taking positive values. A higher value of $\delta$ gives a higher value of the maximum peak in $q_{s}$ while a higher value of $\xi$ displaces this maximum peak to the right.}
	\label{qs}
\end{figure}

\begin{figure}[htbp]
  \centering
  {\includegraphics[width=0.43\textwidth]{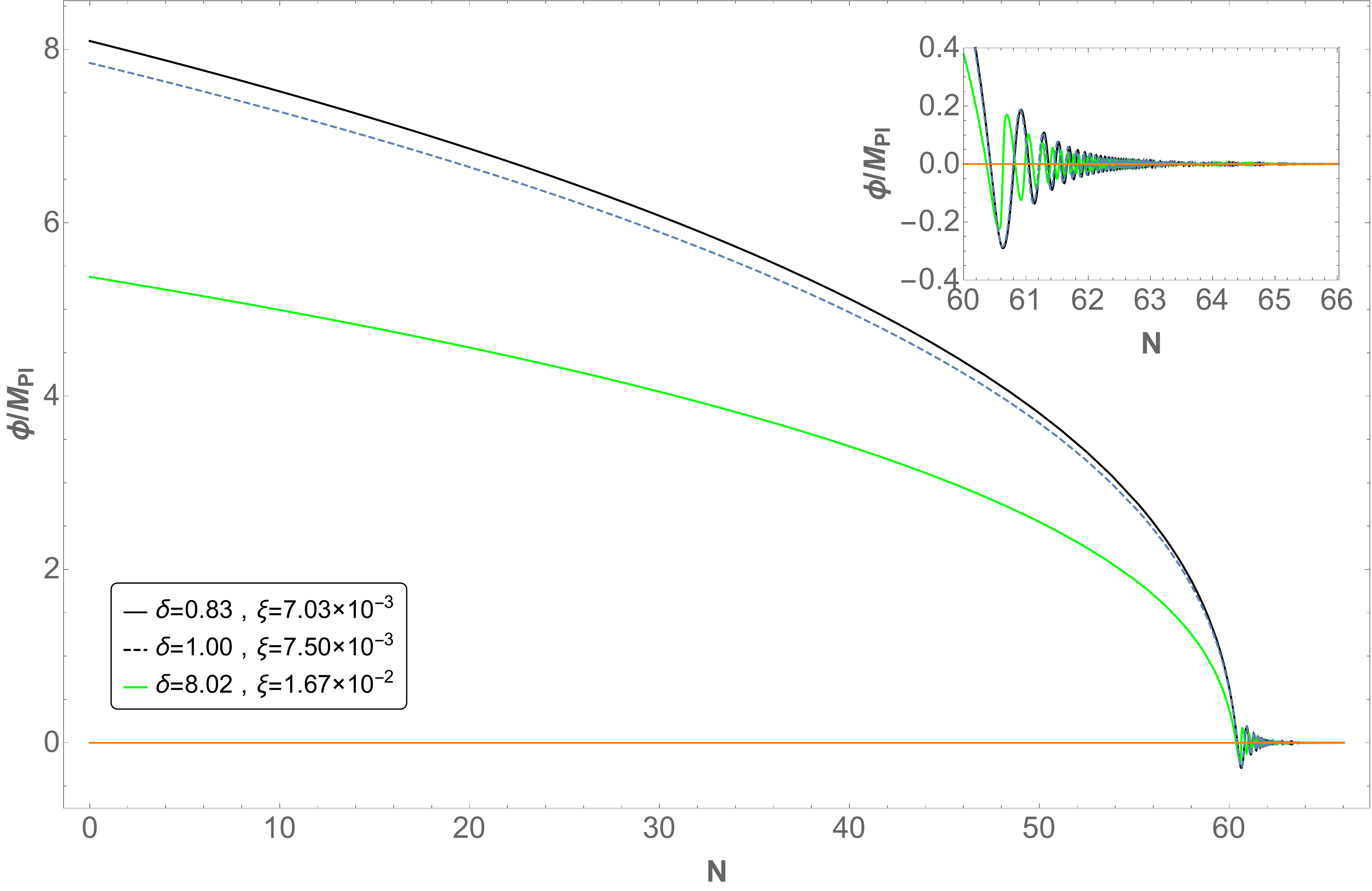}\label{fig:f1}}
  \hfill
 {\includegraphics[width=0.435\textwidth]{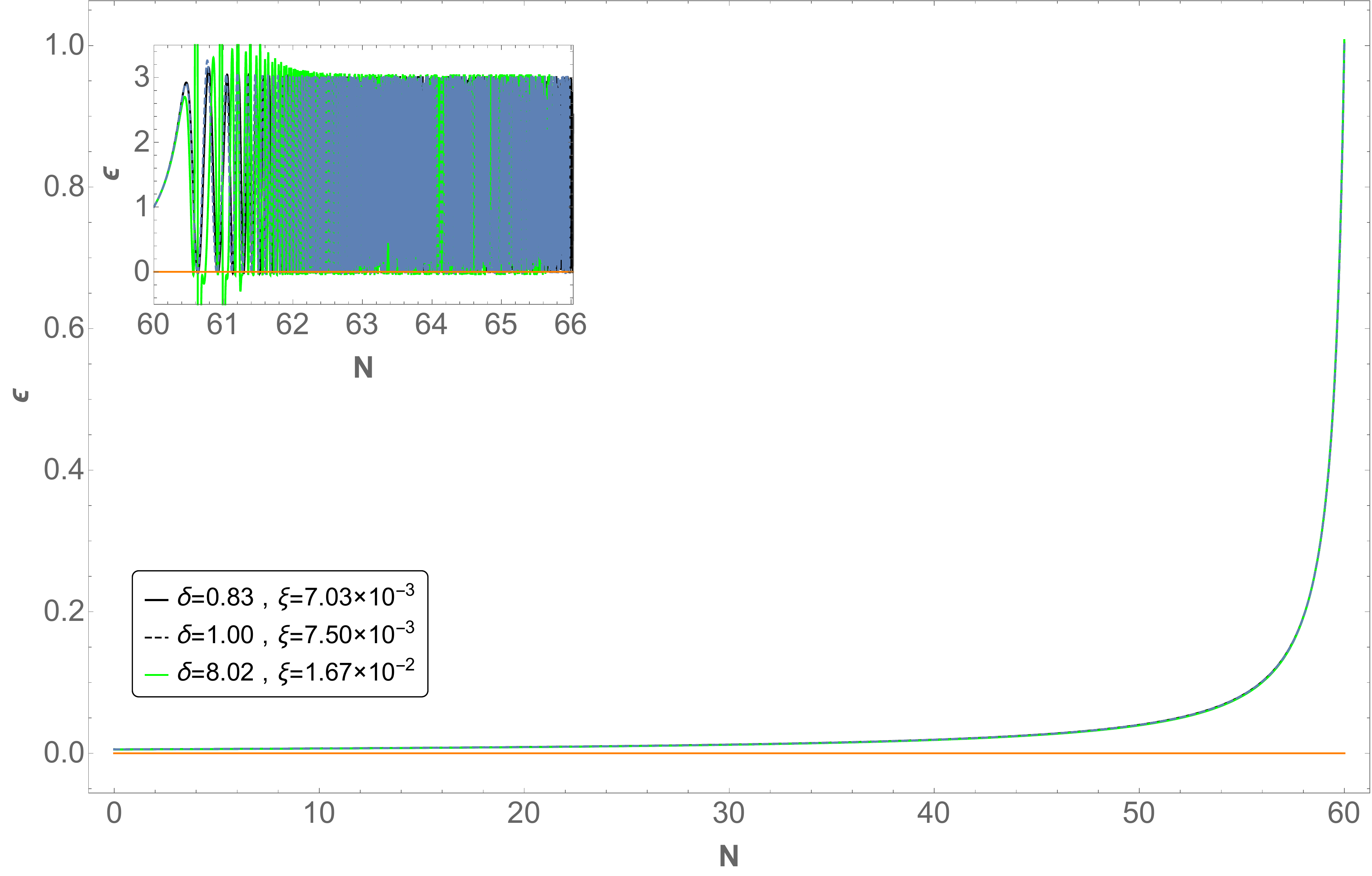}\label{fig:f2}}
  \caption{\small{Left plot: The behavior of the normalized inflaton field for the same values of $\delta$ and $\xi$. One can see that higher values of $\delta$ and $\xi$ implies a lower value of $\phi/M_{Pl}$ at the Hubble horizon crossing. Right plot: During slow-roll inflation the $\epsilon(N)$ function satisfies $\epsilon \ll 1$ and then at the end of inflation, when $N=60$, one verifies that it is equal to $1$. After the end of inflation the $\epsilon(N)$ parameter does not stabilize at any fixed value because we do not assume any coupled matter fluid.}}
	\label{phi_epsilon}
\end{figure}

Since that values of the Galileon self-interaction can affect the oscillatory regime of the inflaton, we can find an upper bound for the parameter $\delta$ by studying the dynamical behavior of scalar propagation speed squared $c_{s}^2$ after the end of inflation \cite{Ohashi:2012wf}. There are two ways to get this upper bound for $\delta$ from the $c_{s}^2$ curve oscillating around $1$ (quadratic potential). The first one consists in to obtain the maximum value for $\delta$ such that the minimum value of $c_{s}^2$ during the post-inflationary epoch is equal to zero, and then avoiding negative values of it. If we strictly avoid $c_{s}^{2}<0$, this means that neither the Galileon term nor the non-minimal coupling term are dominant relative to the standard kinetic term. In this case, for $N=60$, we obtain the upper bound
\be
\delta<0.83<\delta_{c}(\xi),
\ee being that for $\delta=0.83$ we get 
\be
6.95\times 10^3\lesssim \xi\lesssim 1.91\times 10^2.
\ee  On the other hand, the second way is to assume a less stringent limit by allowing larger values of the Galileon self-interaction parameter $\delta$, but for which the inflaton still oscillates coherently during reheating \cite{Ohashi:2012wf}. In this case the Galileon self-interaction dominates on the standard kinetic term even after the end of inflation and $c_{s}^2$ takes negative values. Although it enters in the regime of negative values, it quickly exits this regime due to the oscillations around $1$. So, for $N=60$, we obtain the upper bound
\be
\delta_{c}(\xi)<\delta<\delta_{max}\simeq 8.02,
\ee and for $\delta=\delta_{max}$ we get 
\be 
1.67\times 10^{-2}\lesssim \xi \lesssim 2.88\times 10^{-2}.
\ee In this case, a superluminal behavior can also occur such that $\abs{c_{s}^2}>1$, but this situation does not necessarily lead to a violation of causality \cite{Adams:2006sv,Babichev:2007dw,Bruneton:2006gf,Evslin:2011vh,Evslin:2011rj,Burrage:2011cr}. In FIG \ref{cs2} we have depicted the above mentioned results for $c_{s}^2$ as a function of the $e$-folds number $N$, for different values of the parameters $\xi$ and $\delta$. Moreover, in FIG. \ref{qs}, we have checked that the condition $q_{s}\equiv Q_{s}/M_{Pl}^2 F>0$ is satisfied, which guarantees the  requirement to avoid ghost instabilities \cite{DeFelice:2011uc}. Also, we plot the evolution of the field in FIG. \ref{phi_epsilon} (left panel), and also the slow-roll parameter $\epsilon$ in FIG. \ref{phi_epsilon} (right panel). From these plots we can corroborate that the end of inflation happens exactly in $60$ $e$-folds, and the oscillating scalar field behaves as a barotropic perfect fluid with radiation-like equation of state.

\section{Reheating after inflation}\label{Reheating}

In this section we will show how the Galileon self-interaction and the non-minimal coupling allow us to indirectly predict scenarios for the subsequent post-inflation stage, known as reheating. The comoving Hubble scale $a_{k} H_{k}=c_{s} k$, when the mode with wavenumber $k$ exited the horizon is related to that of the present day $a_{0} H_{0}$ as
\be
\frac{k}{a_{0} H_{0}}=\frac{a_{k}}{a_{end}}\frac{a_{end}}{a_{re}}\frac{a_{re}}{a_{eq}} \frac{a_{eq} H_{eq}}{a_{0} H_{0}} \frac{H_{k}}{H_{eq}},
\label{k_a_H}
\ee where the $k$ label indicates that the quantities are evaluated at the horizon crossing. The other labels correspond to the end of inflation (end), the end of reheating (re), and the radiation-matter equality (eq). By definition, the number of $e$-folds during each stage is given by $e^{N_{k}}=a_{end}/a_{k}$, $e^{N_{re}}=a_{re}/a_{end}$ and $e^{N_{RD}}=a_{eq}/a_{re}$. The expression for the Hubble rate at the horizon crossing $H_{k}$ is modified by
\begin{equation}
    H_{k}=\pi M_{Pl} \sqrt{\frac{\mathcal{P}_{s} r_{k} F_{k}}{2}},
\end{equation} being that we have used Eqs. \eqref{P_t} and \eqref{rr}, with $r_{k}=\left.r\right|_{c_{s} k=a H}$ and $F_{k}=\left.F\right|_{c_{s} k=a H}$.

On the other hand, we assume the matter-energy component of the universe during reheating behaves as a perfect fluid and then its energy density $\rho_{re}$ can be related to the energy density at the end of inflation through the equation \cite{paper2}
\be
\frac{\rho_{re}}{\rho_{end}}=\exp{\left[-3 N_{re} (1+w_{re})\right]}.
\label{Rel}
\ee
In the presence of both the non-minimal coupling to curvature and the Galileon self-interaction one obtains for $\rho_{end}$ the expression
\be
\rho_{end}=\frac{V_{end} \left(1+\lambda_{end}\right)}{F_{end}},
\label{rho_end}
\ee where
\be
\lambda_{end}=\left.\frac{1+2\mathcal{A}-3 \frac{\delta_{F}}{\delta_{X}}}{\frac{3}{\delta_{X}}-1-2\mathcal{A}+3\frac{\delta_{F}}{\delta_{X}}}\right|_{t=t_{end}},
\ee with $V_{end}=V(x_{end})$, $F_{end}=F(x_{end})$, being $x_{end}=\phi(t_{end})/M_{Pl}$. In the absence of both the non-minimal coupling and the Galileon self-interaction, the end of inflation occurs when $\epsilon=\delta_{X}\approx 1$ and then $\lambda_{end}\approx 1/2$.

To calculate the energy density at reheating we follow the standard considerations that consistent in assuming the conservation of entropy in a comoving volume by an appropriate counting of the effective number of relativistic degrees of freedom  $g_{s,re}$ at reheating  and the relation between the current neutrino temperature $T_{\nu,0}$ and the temperature today $T_{0}$ \cite{paper2}. Then one obtains
\be
\rho_{re}=\left(\frac{\pi^2 g_{re}}{30}\right) T_{re}^4=\left(\frac{\pi^2 g_{re}}{30}\right)\left(\frac{43}{11 g_{s,re}}\right)^{4/3}\left(\frac{a_{0}}{a_{eq}}\right)^4\left(\frac{a_{eq}}{a_{re}}\right)^4 T_{0}^{4}.
\label{rho_re}
\ee
Putting Eqs. \eqref{rho_end} and \eqref{rho_re} into Eq. \eqref{Rel} we get $N_{RD}$ as a function of $N_{re}$ and after substituting this result in Eq. \eqref{k_a_H} finally we find
\bea
&& N_{re}=\left(\frac{4}{1-3 w_{re}}\right)\Bigg[-N_{k}-\log\left(\frac{k}{a_{0} T_{0}}\right)-\frac{1}{4}\log\left(\frac{30}{g_{re} \pi^2}\right)-\frac{1}{3}\log\left(\frac{11 g_{s,re}}{43}\right)-\nonumber\\
&& \frac{1}{4}\log\left(\frac{V_{end}}{F_{end}}\right)-\frac{1}{4}\log\left(1+\lambda_{end}\right)+\frac{1}{2}\log\left(\frac{\pi^2 M_{Pl}^2 \mathcal{P}_{s} r_{k} F_{k}}{2}\right)\Bigg],
\eea
where we can assume $g_{s,re}\simeq g_{re}\simeq 100$ and $k=0.002$ $M_{pc}^{-1}$.
Thus, from Eq. \eqref{rho_re} the temperature at reheating is given by
\be
T_{re}=\exp\left[-\frac{3}{4}\left(1+w_{re}\right) N_{re}\right] \left(\frac{3}{10 \pi^2 }\right)^{1/4}\left(1+\lambda_{end}\right)^{1/4}\frac{V_{end}^{1/4}}{F_{end}^{1/4}}.
\ee
Here, the model-dependent expressions are the Hubble rate at the instant when the cosmological scale crosses the Hubble radius, $H_k$, the inflaton potential at the end of the inflationary expansion, $V_{end}$, and the ratio $\lambda_{end}$.
Thus, it is implicit that $N_{re}$, $T_{re}$ depend on the observables ${\mathcal{P}}_s$, $n_s$ and $r$ that we have already discussed.

\subsection{Numerical Results}
 
\begin{figure}[htbp]
	\centering
		\includegraphics[width=0.7\textwidth]{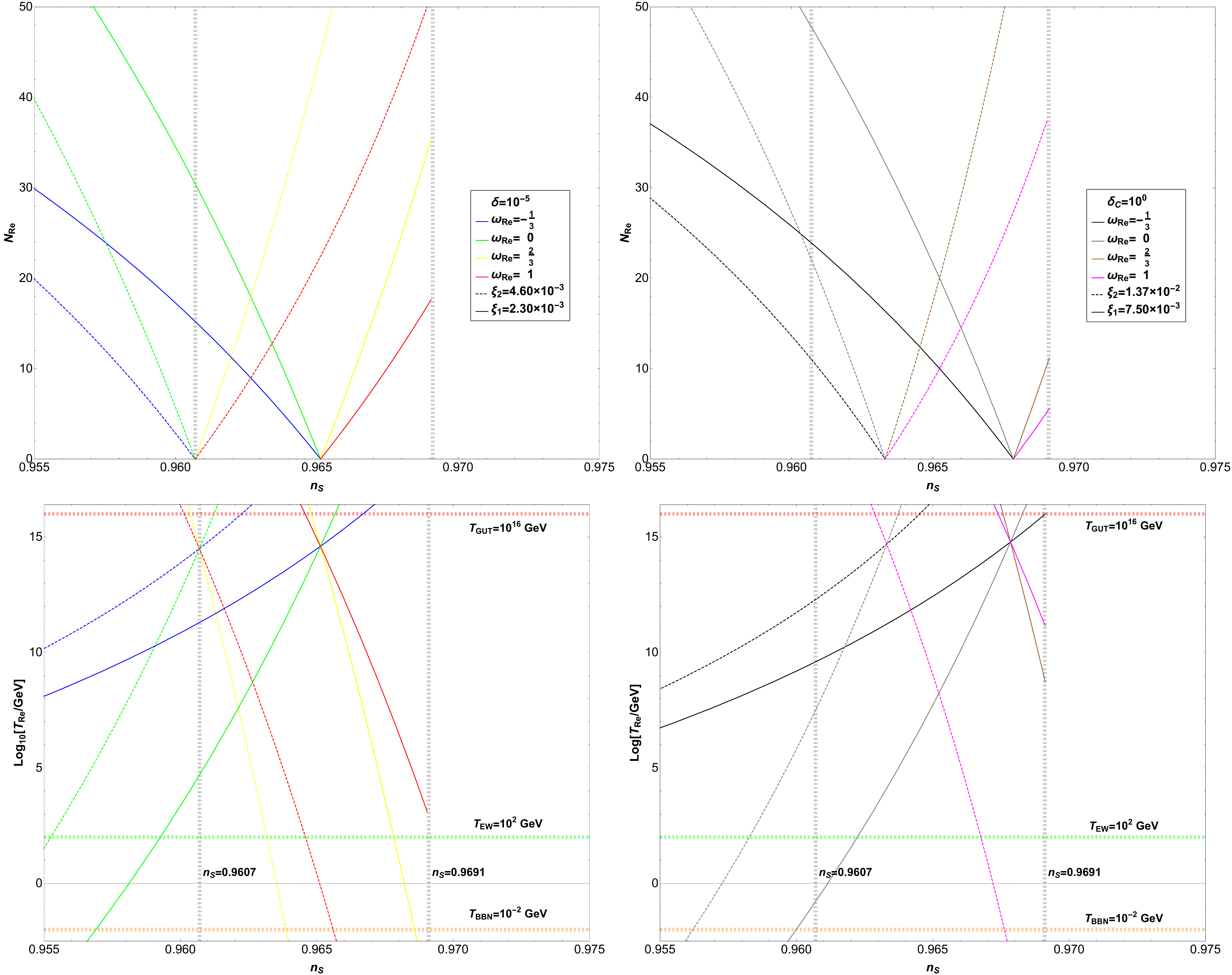}
	\caption{We depict the $e$-folds number of reheating $N_{re}$ and the temperature at the end of reheating $T_{re}$ against the spectral index $n_{s}$. In the left panel, we have chosen the Galileon parameter $\delta=10^{-5}$ and thus the results obtained can be applied to the case of the non-minimally coupled scalar field model. On the other hand, in the right panel we take $\delta=1$ and the effect of the Galileon coupling on $N_{re}$ and $T_{re}$ can be observed. The grey dotted-dashed vertical lines indicate the region associated to the $1\sigma$ bound on $n_{s}$ from PLANCK \cite{planck4}. The orange dotted-dashed horizontal line represents the lower bound on the reheating temperature from BBN.  }
	\label{fig:a}
\end{figure}

We may plot $N_{re}$ and $T_{re}$ versus the scalar spectral index for several different values of the effective equation-of-state parameter $w_{re}$ over the range $-\frac{1}{3}\leq w_{re}\leq1$, as well as for several different values of $\delta$ and $\xi$ that encode the information about the
model. In Fig. \ref{fig:a}, by choosing some values of $w_{re}$, we compare the case when the Galileon parameter $\delta$ is small (left panel) to the case when $\delta$ is equal to the critical value $\delta_{c}=1$ (right panel). For each case, we have considered the respective range of the non-minimal coupling parameter $\xi$ such that the results obtained for $n_{s}$ and $r$ overlap with the $2 \sigma$ C.L. region of the latest PLANCK data. Recall that a small $\delta$ means a high value for the mass scale of the Galileon, because $\delta \sim \frac{1}{M^3}$. We observe that for a fixed value of $\delta$, the instantaneous reheating point ($N_{re}=0$) is shifted to the left, i.e. smaller values of $n_{s}$, when the non-minimal coupling parameter $\xi$ is increasing. Note that the dashed curves in Fig. \ref{fig:a} correspond to the greatest possible value of $\xi$ for each value of $\delta$. On the other hand, when we increase the Galileon parameter, the instantaneous reheating point is now displaced to the right, i.e. higher values of $n_{s}$. Other interesting scenarios take place when we demand the lowest energy scale at which all the Standard Model (SM) particles become ultra relativistic particles. The lowest energy scale to predict such a scenario is considered as the Electroweak stage whose characteristic temperature scale is $T_{EW}=10^{2}$ GeV.  In FIG. \ref{fig:a} we have also included both the constraints on the temperature of reheating coming from the Electroweak scale and the Big Bang nucleosynthesis (BBN) scale $T_{BBN}=10^{-2}$ GeV. Accordingly, we have obtained constraints for the number of $e$-folds of inflation $N_{k}$ that are shown in Table \ref{Table2}. In table \ref{Table2} we summarized our predictions for $w_{re}=-1/3, 0, 2/3, 1$ at the Electroweak scale. We found that by increasing the parameters $\delta$ and $\xi$, the duration of reheating increases too. Finally, in Table \ref{Table2} we also summarized the case when instantaneous reheating takes place. One can observe that instantaneous reheating is reached for $N_{k}\simeq 60$.

\def\tablename{Table}%
\begin{table*}[!ht]
\centering
\begin{center}
\renewcommand{\tabcolsep}{0.5pc} % enlarge column spacing
\renewcommand{\arraystretch}{1} % enlarge line spacing
\begin{tabular}{| c | c | c | c | c | c | c | }\hline\cline{1-7}
$\delta$ &   $\xi \times 10^{2}$  & $N_{re}(w_{re}=-1/3)$ & $ \textup{Log}_{10}\left[\frac{T_{re}}{\textup{GeV}}\right](w_{re}=-1/3)$ & $N_{k}$  & $n_{s}$ & $r$ 
\\\cline{1-7} 
 $10^{-5}$ & $0.230$ & $58.00$ & $2.00$ &  $31.80$ & $ 0.936 $ & $ 0.173 $\\\cline{1-7}
  $1$ & $0.750$ & $58.86$ & $2.00$ &  $30.90$ & $0.940$ & $0.166$\\\cline{1-7}
 $8.02$ & $1.674$ & $58.90$ & $2.00$ &  $30.80$ & $0.940$ & $0.164$\\\cline{1-7}
$\delta$ &   $\xi \times 10^{2}$  & $N_{re}(w_{re}=0)$ & $ \textup{Log}_{10}\left[\frac{T_{re}}{\textup{GeV}}\right](w_{re}=0)$ & $N_{k}$  & $n_{s}$ & $r$ 
\\\cline{1-7} 
 $10^{-5}$ & $0.230$ & $38.66$ & $2.00$ &  $50.84$ & $ 0.959 $ & $ 0.090 $\\\cline{1-7}
  $1$ & $0.750$ & $39.24$ & $2.00$ &  $50.23$ & $0.962$ & $0.088$\\\cline{1-7}
 $8.02$ & $1.674$ & $39.27$ & $2.00$ &  $50.14$ & $0.962$ & $0.087$\\\cline{1-7}
$\delta$ &   $\xi \times 10^{2}$  & $N_{re}(w_{re}=2/3)$ & $ \textup{Log}_{10}\left[\frac{T_{re}}{\textup{GeV}}\right](w_{re}=2/3)$ & $N_{k}$  & $n_{s}$ & $r$ 
\\\cline{1-7} 
 $10^{-5}$ & $0.230$ & $23.18$ & $2.00$ &  $66.11$ & $ 0.967 $ & $ 0.059 $\\\cline{1-7}
  $1$ & $0.750$ & $23.53$ & $2.00$ &  $65.76$ & $0.970$ & $0.060$\\\cline{1-7}
 $8.02$ & $1.674$ & $23.55$ & $2.00$ &  $65.68$ & $0.970$ & $0.059$\\\cline{1-7}
$\delta$ &   $\xi \times 10^{2}$  & $N_{re}(w_{re}=1)$ & $ \textup{Log}_{10}\left[\frac{T_{re}}{\textup{GeV}}\right](w_{re}=1)$ & $N_{k}$  & $n_{s}$ & $r$ 
\\\cline{1-7} 
 $10^{-5}$ & $0.230$ & $19.33$ & $2.00$ &  $69.94$ & $ 0.969 $ & $ 0.054 $\\\cline{1-7}
  $1$ & $0.750$ & $19.62$ & $2.00$ &  $69.65$ & $0.971$ & $0.055$\\\cline{1-7}
 $8.02$ & $1.674$ & $19.63$ & $2.00$ &  $69.57$ & $0.971$ & $0.054$\\\cline{1-7}
$\delta$ &   $\xi \times 10^{2}$  & $N_{re}$ (I.Re.) & $ \textup{Log}_{10}\left[\frac{T_{re}}{\textup{GeV}}\right]$ & $N_{k}$  & $n_{s}$ & $r$ 
\\\cline{1-7} 
 $10^{-5}$ & $0.230$ & $0$ & $14.60$ &  $60.38$ & $ 0.965 $ & $ 0.069 $\\\cline{1-7}
  $1$ & $0.750$ & $0$ & $14.80$ &  $59.94$ & $0.967$ & $0.067$\\\cline{1-7}
 $8.02$ & $1.674$ & $0$ & $14.70$ &  $59.85$ & $0.970$ & $0.067$\\\cline{1-7}
 \hline
\end{tabular}
\end{center}
\caption{Summary of the reheating predictions of our model for several different values of $w_{re}$ at the Electroweak scale, including the duration of reheating $N_{re}$, the temperature of reheating $T_{re}$, the number of $e$-folds of inflation $N_{k}$ and the inflationary observables $n_{s}$ and $r$. Also, we have included the results for the scenario of instantaneous reheating (I. Re.)} 
\label{Table2}
\end{table*}

\begin{figure}[htbp]
  \centering
  \includegraphics[width=0.7\textwidth]{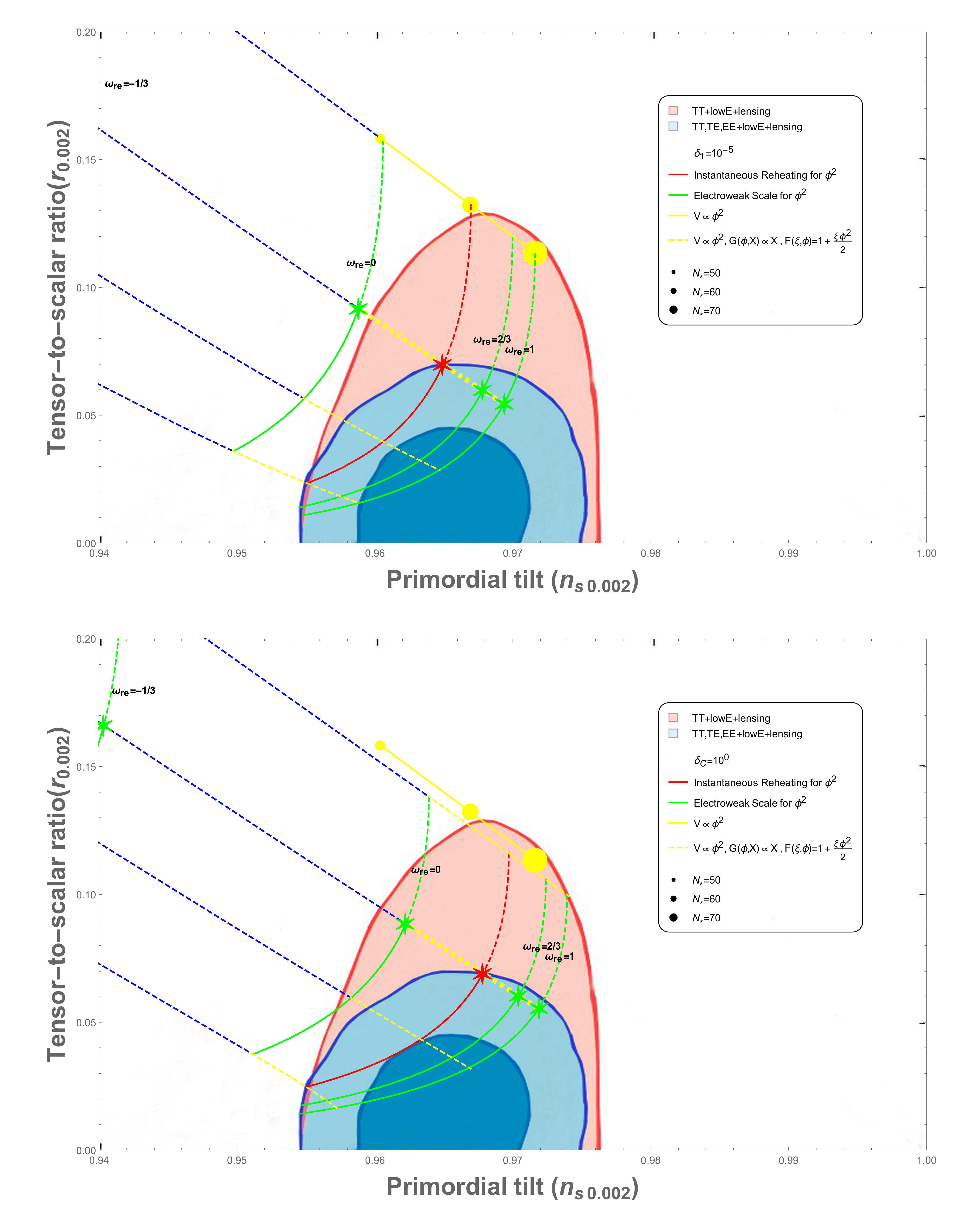}
  \caption{We show the $n_{s}-r$ plane along with the latest constraints from PLANCK and the results obtained from reheating for the model at hand. The dashed red curve indicates instantaneous reheating while the dashed green line the Electroweak scale. Several different values of the equation of state of reheating have been represented.}
		\label{fig:b}
\end{figure}

In Fig. \ref{fig:b} we plot again the predictions of our model in the $n_{s}-r$ plane but now taking into account the new constraints obtained for the $e$-folds number of inflation $N_{k}$ from the reheating scenario, and then the several different values of $w_{re}$. In the left panel of Fig. \ref{fig:b} we take a small value for the Galileon parameter, $\delta=10^{-5}$, and thus the results obtained are very close to the case of a non-minimally coupled scalar field model \cite{paper2}. On the other hand, for the right panel of Fig. \ref{fig:b}, we consider the case when the Galileon self-interaction parameter assumes its critical value, i.e. $\delta_{c}=1$. The instantaneously reheating is denoted by the red line with the red star indicating the point where the results enter into the $2 \sigma$ C.L. region of PLANCK. The Electroweak scale is predicted differently according to the value of $w_{re}$ and the duration of inflation as shown in Fig. \ref{fig:b} through the green lines. Also, in this plot the green star corresponds to the intersection between the straight line associated to the lowest value of $\xi$ (Fig. \ref{fig:a}) and the curves obtained from the Electroweak temperature bound. In Fig. \ref{fig:b} we observe that for $w_{re}$ closer to $w_{re}=1$ it is required a higher value of $N_{k}$, in order to achieve the Electroweak scenario. %This means that the last perturbations produced by the chaotic potential in the inflation stage were produced more closely to the singularity of the big-bang (as $N_{k}$ grows). 
Interestingly, both cases of instantaneous reheating for the chaotic potential in the scalar-tensor theory with Galileon self-interaction and non-minimal coupling require an inflationary stage whose duration was at least $N_{k}\simeq 60$ $e$-folds. Moreover, comparing the left panel ($\delta=10^{-5}$) and the right panel ($\delta=\delta_{c}=1$) in Fig. \ref{fig:b}, we note that a higher value of $\delta$ leads to a displacement of all the predictions to the right for higher values of $n_{s}$. Finally, from this plot we obtain new constraints for the non-minimal coupling parameter by using the constraints already obtained for $N_{k}$ from reheating (Table \ref{Table2}). For example, for $\delta=10^{-5}$ and $w_{re}=2/3$ we find  $1.81\times 10^{-3}\lesssim \xi \lesssim 7.95 \times 10^{-3}$, while for $w_{re}=1$ one has $1.7\times 10^{-3}\lesssim \xi\lesssim 8.34 \times 10^{-3}$. Similarly, for $\delta=\delta_{c}=1$ and $w_{re}=2/3$, we find $6.9\times 10^{-3}\lesssim \xi\lesssim 2.13 \times 10^{-2}$, while for $w_{re}=1$ the allowed range for $\xi$ is  $6.9\times 10^{-3} \lesssim \xi \lesssim 2.16 \times 10^{-2}$.

\begin{figure}[htbp]
	\centering
		\includegraphics[width=0.5\textwidth]{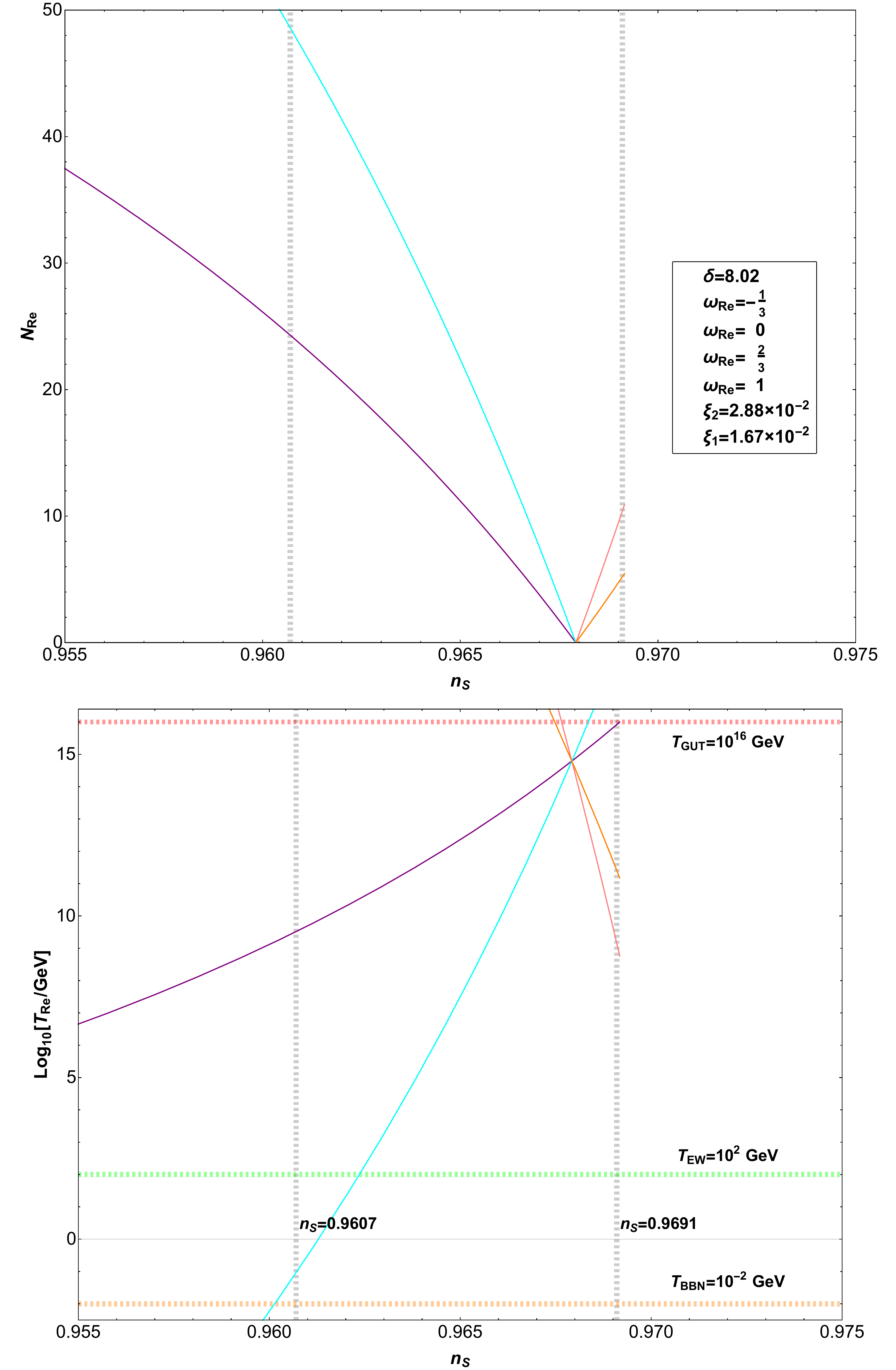}
	\caption{We depict the $e$-folds number of reheating $N_{re}$ and the temperature at the end of reheating $T_{re}$ as a functions of the spectral index $n_{s}$. For the Galileon parameter we take the critical value $\delta=8.02$. This is the highest value of $\delta$ for which  we find coherent oscillations of the inflaton after the end of inflation.}
	\label{fig:2}
\end{figure}

\begin{figure}[htbp]
	\centering
		\includegraphics[width=0.7\textwidth]{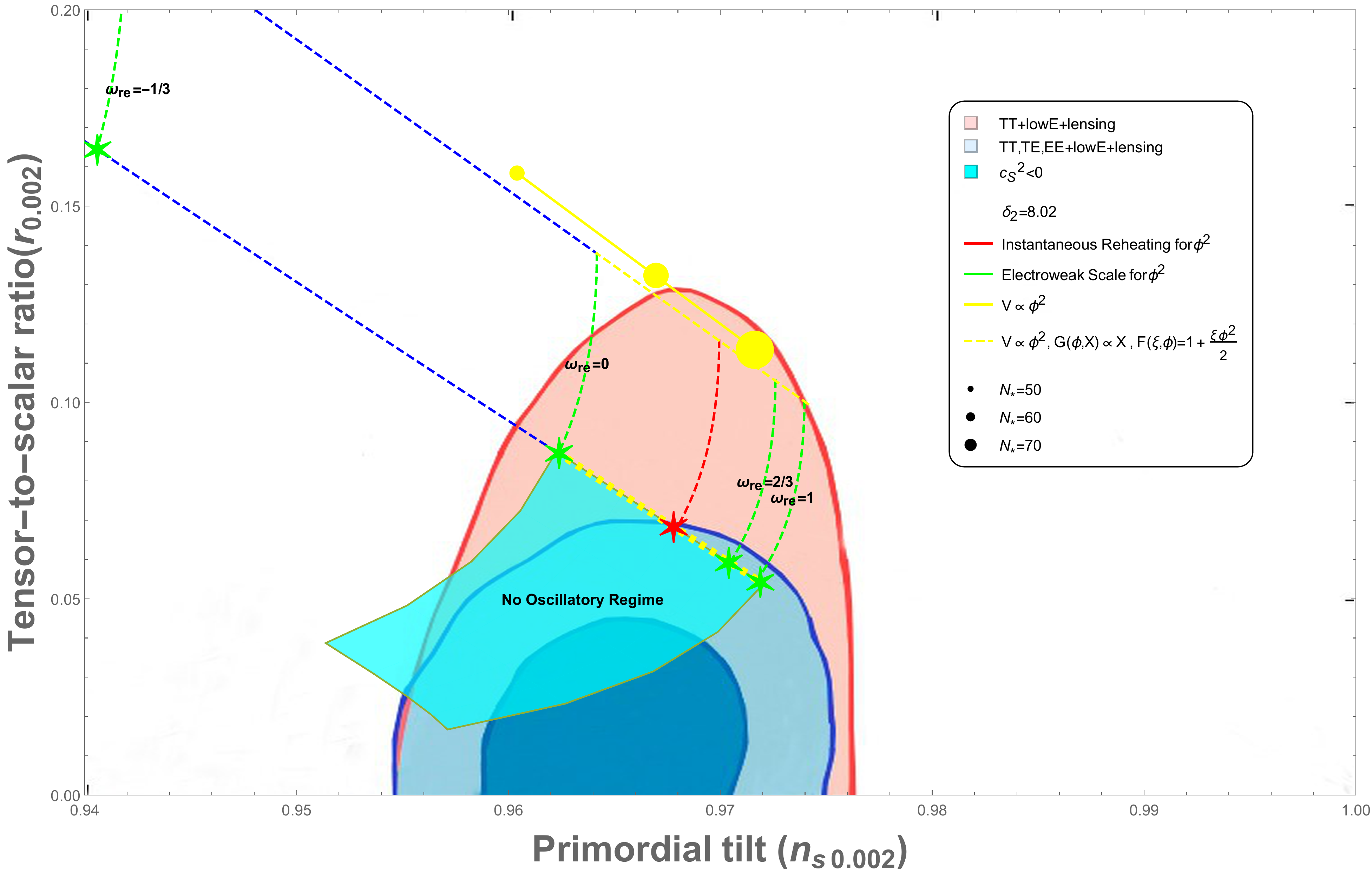}
	\caption{The $n_{s}-r$ plane for the highest value of Galileon parameter $\delta$ that allows for coherent oscillations of the inflaton after the end of inflation. It is observed that some region of the parameter space could lead to a no oscillatory regime and then to an impossibility of the reheating stage.  }
	\label{fig:c}
\end{figure}

In Fig. \ref{fig:2}, we summarize the results obtained for the maximum value of the Galileon self-interaction parameter $\delta$ allowing coherent oscillations after the end of inflation. In the reheating scenario, with this value instantaneously reheating was predicted just in the highest value predicted for $n_{s}$ by our model. In Fig. \ref{fig:c} one can observe that instantaneous reheating is still achieved at $N_{k}\simeq 60$. However, by evaluating numerically the field equations we observed that our model does not satisfied the upper bound of $r \leq 0.064$ from PLANCK data. 
Thus, in this case an instantaneous reheating does not give a reliable prediction. On the other hand, by observing the results obtained for the Electroweak scale with $w_{re}=2/3$ and $w_{re}=1$, the predictions are consistent with the upper bound on $r$ set by PLANCK.

%%%%%%%%%%%%%%%%%%%%%%%%%%%%%%%%%%%%%%%%%%%%%%%%%%%%%%%%%%%%
\section{Concluding Remarks}\label{Concluding_Remarks}

We studied slow-roll inflation and reheating in non-minimally coupled scalar-tensor gravity theory with Galileon self-interaction. This theory corresponds to a wide sector of the surviving Horndeski theory (without including k-essence) after the recent constraints from Gravitational Waves experiments \cite{TheLIGOScientific:2017qsa,Baker:2017hug}. Thus, we extended the results obtained in Ref. \cite{paperbase} for the potential-driven Galileon model by including a non-minimal coupling between the scalar field and the curvature scalar.

In Ref.~\cite{Weinberg:2008hq}, Weinberg used the methods of effective field theory (EFT) to study generic theories of inflation with a single inflaton field. He started from the action of a minimally coupled scalar field with a generic potential $V(\phi)$, since a possible non-minimal coupling to gravity can be removed through a conformal transformation of the spacetime metric and field redefinitions \cite{Clifton:2011jh}. Then, by expanding in inverse powers of some large mass $M$, he established the leading correction to this action as given by the sum of all the possible generally covariant terms constructed with four spacetime derivatives, and coefficients that depend on the scalar field. Regarding this, all the second time derivatives, as it is the case of the Galileon self-interaction term $G(\phi,X)\Box{\phi} \sim f(\phi) X \Box{\phi}$ and the time derivatives of the auxiliary fields, can be eliminated by using the zeroth order field equations. Note that this approach to the effective field theory for inflation is different from the one followed in Ref.~\cite{Cheung:2007st} by Cheung et al. By starting from the unitary gauge they studied the most general theory with broken time diffeomorphisms that describes the fluctuations around a quasi de Sitter background in single-field inflation models. 

In the present paper we studied inflation in single-field models from the view of point of the Horndeski theory \cite{Horndeski:1974wa,Kobayashi:2019hrl}. Particularly, we started from the action \eqref{Action1} which includes two important sectors of the Horndeski theory, the non-minimal coupling of the scalar field to gravity and the Galileon self-interaction \cite{Kobayashi:2019hrl}. A key aspect of this approach is that in general one comes across in the field equations with a kinetic mixing of gravity and the scalar field that does not happen in the case of Einstein gravity plus a minimally coupled scalar field. Of course, for the case in which we only have the non-minimal coupling to gravity this mixing can be undone under a conformal transformation of the spacetime metric and field redefinitions as previously explained. However, in the presence of both the non-minimal coupling and the Galileon self-interaction this mixing becomes essential and there is not a way to remove it \cite{Kobayashi:2019hrl}. This important feature constitutes the so-called kinetic gravity braiding in Horndeski theory \cite{Pujolas:2011he}. Finally, it is also worth noting that there is a connection between the Horndeski theory and the EFT formalism of Cheung et al. The Horndeski theory belongs to a sub-class of the corresponding general EFT action \cite{Gleyzes:2013ooa,Bloomfield:2013efa,Tsujikawa:2014mba}. 

Thus, after working out the expressions for the power spectra of scalar and tensor perturbations, we studied the predictions for chaotic inflation in the presence of both the non-minimal coupling to the curvature scalar and the Galileon self-interaction. We showed the transition point $x=x_{G}\equiv \phi_{G}/M_{Pl}$ from the Galileon-dominated regime to the standard inflation is modified by the presence of the non-minimal coupling to curvature. Hence, we studied the implications of this in terms of the corresponding critical value of the Galileon self-interaction parameter $\delta=\delta_{c}(\xi)$ related to the transition point which becomes a function of the non-minimal coupling parameter $\xi$. Then, we confronted the predictions of the model in the $n_{s}-r$ plane by using the latest PLANCK data \cite{planck4}. Particularly, we showed the Galileon self-interaction parameter plays an important role in lowering the predictions for the tensor-to-scalar ratio parameter $r$ in the non-minimally coupled scalar-tensor theories.  Therefore, the combined effect of the Galileon self-interaction and the non-minimal coupling to curvature allows predictions for quadratic inflation compatible not only with the $95\%$ C.L region but also slightly inside the $68\%$ C.L region (see Fig. \ref{r_ns_compressed}). In Table \ref{Table1} we summarized all our results for the constraints on the parameters of the model, including the constraints on the mass scale of the inflaton field $m_{\phi}$ and the Galileon mass $M$.

Although the Galileon self-interaction of the scalar field provides a viable mechanism to significantly reduce the tensor-to-scalar ratio parameter $r$ in the non-minimally coupled scalar-tensor theories, the corresponding interaction parameter $\delta$ cannot be upper constrained by solely using the  $n_{s}-r$ plane. To overcome this issue one could for instance resort to the framework of effective field theories by considering slow-roll inflation as a low-energy effective theory with the UV cut-off given by the Planck scale, $\Lambda\lesssim M_{pl}$, and the minimal cut-off determined by the inflationary Hubble scale, $\Lambda \simeq M \gtrsim H$, with the masses of the light fields satisfying $m_{\phi}\lesssim H$ \cite{Baumann:2014nda}. However, in order to obtain results that are compatible with the subsequent evolution stage that is reheating we can constrain the Galileon parameter $\delta$ by studying the oscillatory regime of the scalar after the end of inflation \cite{paperbase}. To carry out this study we have numerically solved the complete equations of the model by introducing a set of appropriated dimensionless variables that parameterize the phase space of the inflaton field. Then we obtained two kind of upper bound for the parameter $\delta$ that extend the results obtained in Ref. \cite{paperbase} to the case when the non-minimal coupling is switched on. The first kind of upper bound obtained for $\delta$ is consistent with non-negative values of the scalar propagation speed squared, $c_{s}^{2}>0$, and the dominance regime of the Galileon self-interaction term ends before the end of inflation. On the other hand, the second kind of upper bound obtained for $\delta$ is a less stringent limit by allowing negatives values of the scalar propagation speed squared, $c_{s}^{2}<0$, but for which the inflaton still oscillates coherently during reheating. In this latter case the dominance of the Galileon term is slightly extended until after the end of inflation but without spoiling reheating. Then we found that the parameter space associated with the non-minimal coupling $\xi$ is also constrained by the condition for the fulfilment of the oscillatory regime of the inflaton.

Finally, we studied reheating after inflation by using the perfect fluid approximation for which reheating is parameterized in terms of the equation of state (EOS) parameter $-1\leq w_{re} \leq 1$ \cite{reh1, paper1}. Then, we got the relations between the duration of reheating $N_{re}$, the temperature at the end of reheating $T_{re}$, the equation of state $w_{re}$ and the number of $e$-folds of inflation $N_{k}$. These relations are model dependent and they allow us to connect the reheating predictions with the inflationary observables and then the current PLANCK data for inflation \cite{paper2, paper3}. Particularly, by using the physical requirements from the Big Bang Nucleosynthesis (BBN) and the Electroweak (EW) scale for the temperature of reheating, along with the constraints for the spectral index $n_{s}$ from PLANCK data, we obtained fine-tuned constraints on the duration of slow-roll inflation $N_{k}$ (see Figs. \ref{fig:a} and \ref{fig:2}) which are shown in Table \ref{Table2}. Then, using these results for $N_{k}$ from reheating we returned to the $n_{s}-r$ plane to obtain improved constraints on the non-minimal coupling parameter $\xi$. In Figs. \ref{fig:b} and \ref{fig:c}, we showed that in the present model the stiff EOS parameter during reheating ($w_{re}>1/3$) is favored by the latest PLANCK data compared to the case of a soft EOS parameter ($w_{re}<1/3$). The stiff EOS parameter yields values for the tensor-to-scalar ratio $r$ which are in agreement with the current upper bound on $r$ reaching a deeper region within the $1 \sigma$ C.L. region than in the case of a soft EOS parameter, and then predicting a broader range for the reheating temperature from the EW scale to the instantaneous reheating below the GUT energy scale.

%%%%%%%%%%%%%%%%%%%%%%%%%%%%%%%%%%%%%%%%%%%%%%%

%\balance
%\clearpage

%%%%%%%%%%%%%%%%%%%%%%%%%%%%%%%%%%%%%%%%%%%%%%%%%%%%%%%%%%%%%%%%%%%%
\section{Acknowledgments}
M. L\'opez acknowledges support from PUCV through Beca de T\'ermino de Tesis. G. Otalora acknowldeges DI-VRIEA for financial support through Proyecto Postdoctorado $2020$ VRIEA-PUCV. N.~Videla was supported by Comisi\'on Nacional de Ciencias y Tecnolog\'ia of Chile through FONDECYT Grant N$^{\textup{o}}$ 11170162.
%%%%%%%%%%%%%%%%%%%%%%%%%%%%%%%%%%%%%%%%%%%%%%%%%%%%%%%%%%%%%%%%%%%%

%\clearpage
%\appendix


%\clearpage

%\bibliography{bio}   % name your BibTeX data base

\end{document}